\documentstyle[epsfig]{mn}

\def\bm{\mathbf} \def\planck{{\sc Planck}} \def\plancks{{\sc Planck }}

\begin{document}
\title{Multi-frequency Wiener filtering of CMB data with polarisation}

\author[Bouchet et al.]  { F. R. Bouchet$^1$, S. Prunet$^{1,2}$, and Shiv
  K. Sethi$^1$ \\ $^{1}$, Institut d'Astrophysique de Paris, CNRS, 98bis
  Boulevard Arago F--75014 Paris France\\ $^{2}$ Institut d'Astrophysique
  Spatiale, B\^at. $121$, 91405 ORSAY, France}

\maketitle

\begin{abstract}

One goal of CMB data analysis is to combine data at different frequencies,
angular resolutions, and noise levels in order to best extract the component
with a Plankian spectral behaviour. A multi-frequency Wiener filtering method
has been proposed in this context by Bouchet, Gispert and Puget (1995) and in
parallel  by Tegmark and Efstathiou (1996). As shown in Bouchet and Gispert
(1998a), this linear method is also convenient to estimate a priori, given a
sky model and an experimental description, the residual errors on the CMB
power spectrum assuming the foregrounds have been removed with this method. In
this paper, we extend the method to the case when additional polarisation data
is available. In particular, we derive the errors on the power spectra
involving polarisation and show numerical results for the specifications of
the future CMB space missions MAP and \plancks\hspace{-2pt}\footnote{For current
  noise specifications and  angular and frequency  coverage of these experiments, see
  {\tt http://map.gsfc.nasa.gov} and {\tt http://astro.estec.esa.nl/SA-general/Projects/Planck}}   when it is assumed that the
Galactic synchrotron and dust emission are respectively about 40\% and 10 \%
polarised.  We consider two  underlying models for our study:
we take a standard CDM model with
$\tau = 0.1$ for  the extraction of $E$-mode
polarisation and $ET$ cross-correlation ; for $B$-mode polarisation we
consider a  tilted CDM  model with $n_s = 0.9$, $n_T = -0.1$ and  $T/S
= 0.7$. We find that:   (1) The
resulting fractional errors on $E$ mode polarisation and $TE$
cross-correlation power spectra are $\la 10 \hbox{--} 30 \, \%$ for $
50 \la  \ell \la  1000$ for \planck. The fractional errors are between
$50 \%$ to $150 \%$ for $\ell \le 50$, (2) The corresponding fractional
errors for MAP are $\ge 300 \, \%$ for most of the $\ell$ range,
(3)   the Wiener filtering give extraction errors $\le 2$
times  the expected
performance for the  combined sensitivity of all the channels of
\plancks. For MAP, the corresponding degradation is $\simeq 4$.
(4) if, instead of individual modes, one considers band-power
estimates with a logarithmic interval
$\Delta \ell /\ell = 0.2$ then  the fractional error for MAP
drops to   $\la 100 \%$  at the Doppler peak around $\ell \simeq
300$ for the  $ET$ signal,
 and (5) The fractional error for 
 $B$-mode polarisation detection is  $ \la 100 \%$
with \plancks for $\ell \le 100$.
A band-power estimate with  $ \Delta \ell /\ell = 0.2$
 reduces the fractional
errors  to $\la 25 \%$ for $20 \le \ell \le 100$.

\end{abstract}

\section{Introduction}

Ever since the detection of CMB temperature anisotropies by the DMR experiment
aboard the COBE satellite at angular scales $\ge 7^{\circ}$ (Smoot {\em et al.
\/} 1992, Bennett {\em et al. \/} 1996), there has been a surge of activity on
both theoretical and experimental front. Several detections at smaller angular
scales have been reported (for a recent compendium, see Lineweaver and Barbosa
1998). There is great interest in the upcoming satellite projects MAP and
\planck. These all-sky experiments, it is hoped, will determine the angular
power spectrum of CMB temperature fluctuations at all scales greater than a
few arc minutes. Various theoretical analyses have shown under fairly general
assumptions that this could determine cosmological parameters, $\Omega$,
$\Omega_B$, $h$, etc, with unprecedented precision (Jungman {\em et al. \/}
1996, Zaldarriaga {\em et al.\/} 1997, Bond {\em et al. \/} 1997). In
addition, there is growing optimism that these satellite projects might detect
the very small expected signal from the polarised component of the CMB. The
information from CMB polarisation could then be combined with temperature
signal 1) to check the self-consistency of the predictions of the underlying
theoretical model, 2) to further break the degeneracy between some
cosmological parameters and 3) to help in unambiguously detecting a
tensor-induced component of CMB fluctuations (Seljak \&  Zaldarriaga 1997,
Seljak 1997, Kamionkowski \& Kosowski 1997).

A major hurdle in extracting the primary CMB signal from data, apart from
noise, is the presence of poorly known Galactic and extra-galactic
foregrounds. However, as the foregrounds differ from the CMB emission in both
frequency dependence and spatial distribution, one can reduce their level by
proper combination of the multi-frequency data of a CMB experiment. Bouchet
{\em et al. \/} (1995) and Tegmark \& Efstathiou (1996) proposed a particular
linear scheme, based on the traditional Wiener filtering method to take full
advantage of this fact. The performance of the method has been assessed
through detailed numerical simulations performed in the context of the
\plancks preparation (Gispert \& Bouchet 1996, Bouchet \& Gispert 1998b).  It
was shown that the residual contamination after cleaning the map is much
smaller than the CMB primary signal, and therefore the foregrounds may not be
a major obstacle in the extraction of CMB temperature angular power spectrum.

The CMB polarisation signal is expected to be one to two order of magnitudes
below the temperature signal. And it is likely to be comparable to the
achievable experimental noise in the current experiments. The presence of
foregrounds should, of course, make this detection even more difficult. In our
Galaxy, the synchrotron emission is highly polarised and would constitute a
major foreground for measurements made in the Rayleigh-Jeans part of the CMB
spectrum (e.g. MAP and Low-Frequency Instrument -LFI- of \planck). The
Galactic dust emission is also observed to be polarised and it is  the
dominant foreground in the Wien part of the CMB spectrum (case of
High-Frequency Instrument -HFI- of \planck). Prunet {\em et al. \/} (1998)
modelled and estimated the level of dust polarised emission at high Galactic
latitudes, and compared it to the polarised component of the CMB signal. They
showed that the scalar-induced polarisation (E-mode) power spectrum and
temperature-polarisation cross-correlation are likely to be much larger than
the dust polarised emission at high Galactic latitudes.  However, the
tensor-induced (B-mode) signal is at best comparable to the foreground
contamination level. The extraction of such a signal in the presence of
comparable instrumental noise, even with small foregrounds, is trickier than
the corresponding temperature case, where the signal is much bigger than the
instrumental noise.

In this paper, we extend the multi-frequency Wiener filtering method to
include the polarisation and temperature-polarisation cross-correlation. The
goal of this exercise is to quantify errors in estimating various power
spectra when Galactic polarised foregrounds are present.
We describe the method in detail in next section. In \S~3, we give a
description of the polarised foregrounds emission which is used in \S~4  to
give the resultant errors on the power spectra estimates in the case of MAP
and \planck.  We summarize our results and discuss their limitations and
applicability in \S~5.

\section{Multi-frequency Wiener filtering on CMB data}

We present here an extension of the multi-frequency Wiener filtering technique
applied earlier on CMB temperature data (Bouchet {\em et al. \/} 1995,
Tegmark \& Efstathiou 1996, Bouchet \& Gispert 1998a, hereafter BG98).
The goal of this
extension is to see the effect of including polarisation data.

Let us denote the observed data at different frequencies as, $y_{\nu}^i$,
where $\nu$ indicates the frequency of the instrumental channel, and $i$ the
nature of the observed field (temperature or $E$-mode polarisation. The $B$
mode polarisation has a vanishing cross-correlation with both $T$ and $E$ and
therefore can be treated separately). The $E$ and $B$ modes are the
'divergence' and 'curl' part of the polarisation tensor and are linear
combinations of the usual Stoke's parameters $Q$ and $U$. The advantage of
using these variables is that the $B$ mode polarisation vanishes for
scalar-induced perturbations (for details see Seljak 1997). The data points
$y_{\nu}^i$, for a given $i$, take contributions from various Galactic and
extra-galactic sources, apart from the primary CMB signal and instrumental
noise.  Throughout this paper, we adopt the convention of Tegmark and
Efstathiou (1996) and take $y_{\nu}^i$ to mean the 'observed' temperature
fluctuation, i.e., the monopole part is first subtracted from the observed
surface brightness, which is then divided by $\partial B_0/\partial T_0$,
$B_0$ being the surface brightness of CMB at $T=T_0$.  Let us call $x_p^j$ the
contribution of the field $j$ due to process $p$, this is the quantity we want
to recover from the observational data $y_{\nu}^i$.  We assume there is a
linear relation between them:
\begin{equation}
y_{\nu}^i=A_{\nu p}^{ij}*x_p^j + b_{\nu}^i ,
\label{fil}
\end{equation}
where $A_{\nu p}^{ij}$ is the instrument response kernel, and $b_{\nu}^i$ is
the detector noise level per pixel for the full mission time (all repeated
indices are summed over). As discussed in BG98, it is easier to deal with the
spherical harmonics transform of Eq.~(\ref{fil}), as this non-local equation
in pixel space transforms into a algebraic linear equation in multipole space.
From now on, all spatially dependent variables are then expressed in multipole
space. Eq.~(\ref{fil}) translates into:
\begin{equation}
y_{\nu}^i (l,m)=A_{\nu p}^{ij} (l,m) x_p^j (l,m) + b_{\nu}^i (l,m)
\label{filmod}
\end{equation}
The problem now is to construct an optimal estimator $\hat x_p^j$. We chose
this estimator to be linear in the observed data:
\begin{equation}
\hat x_p^i = W_{p\nu}^{ij} y_{\nu}^j .
\label{recon}
\end{equation}
The reconstruction error for a given field and process $(\varepsilon_p^i)^2
 \equiv |(\hat x_p^i - x_p^i)^2| $ can be written as:
\begin{eqnarray}
\left(\varepsilon_p^i\right)^2 & = & \left(W_{p \nu}^{ij} A_{\nu p'}^{jk} -
\delta_{pp'} \delta_{ik}\right)\left(W_{p \nu '}^{il} A_{\nu ' p''}^{lm} -
\delta_{pp''} \delta_{im} \right)\langle x_{p'}^k x_{p''}^m\rangle \nonumber
\\ & & + W_{p\nu}^{ia} W_{p\nu '}^{ib} \langle b_{\nu}^a b_{\nu '}^b\rangle ;
\label{reconer}
\end{eqnarray} 
$W_{p\nu}^{ij}$ is chosen so as to make the the variance of the reconstruction
 error minimal.  The derivatives of the error with respect to the filters
 coefficients $W_{p\mu}^{ic}$ should then be zero. This condition can be
 expressed as:
\begin{eqnarray}
& & \left[\left(\delta_{jc}\delta_{\mu\nu}A_{\nu p'}^{jk}\right) \left(W_{p\nu
'}^{il}A_{\nu 'p''}^{lm} - \delta_{pp''}\delta_{im}\right) \right . \nonumber
\\ & & \left . {} + \left(\delta_{\mu\nu '}\delta_{lc}A_{\nu 'p''}^{lm}\right)
\left(W_{p\nu}^{ij}A_{\nu p'}^{jk} - \delta_{pp'}\delta_{ik}\right)\right]
\langle x_{p'}^k x_{p''}^m\rangle \nonumber \\ & & {}+
\left(\delta_{\mu\nu}\delta_{ca}W_{p\nu '}^{ib} + \delta_{\mu\nu
'}\delta_{cb}W_{p\nu}^{ia}\right) \langle b_{\nu}^a b_{\nu '}^b\rangle = 0 .
\end{eqnarray}
Rearranging terms we get the final equation on the filters:
\begin{equation}
A_{\mu p'}^{ck}W_{p\nu '}^{il}A_{\nu 'p''}^{lm}\langle x_{p'}^k
x_{p''}^m\rangle + W_{p\nu '}^{ib}\langle b_{\mu}^c b_{\nu '}^b\rangle =
A_{\mu p'}^{ck}\langle x_{p'}^k x_p^i\rangle .
\label{wien}
\end{equation}

\subsection{Implementation of foregrounds removal}

Eq.~(\ref{wien}) is valid for the general case in which various processes,
fields, and corresponding instrumental noises could be correlated.  We
consider here only uncorrelated processes and uncorrelated noises between
different fields and channels. We allow for the correlation between the two
fields $T$ and $E$. These conditions can be expressed as:

\begin {eqnarray}
\langle x_{p'}^1 x_{p''}^1\rangle & = & \delta_{p'p''}C_{p'}^T  = 
{\bm\Delta}^T_{p'p''} \nonumber \\ \langle x_{p'}^1 x_{p''}^2\rangle & = &
\delta_{p'p''}C_{p'}^{TE} = {\bm\Delta}^{TE}_{p'p''} \nonumber \\ \langle
x_{p'}^2 x_{p''}^2\rangle & = & \delta_{p'p''}C_{p'}^E  =
{\bm\Delta}^E_{p'p''} 
\label{nocross}
\end{eqnarray}
\begin{eqnarray}
\langle b_{\nu '}^1 b_{\nu}^1\rangle & = & \delta_{\nu '\nu}B_{\nu '}^T = {\bm
  N}_{\nu '\nu}^T \nonumber \\ \langle b_{\nu '}^2 b_{\nu}^2\rangle & = &
  \delta_{\nu '\nu}B_{\nu '}^E = {\bm N}_{\nu '\nu}^E \nonumber \\ \langle
  b_{\nu '}^1 b_{\nu}^2\rangle & = & 0
\label{cross}
\end{eqnarray}
 With these conditions, Eq.~(\ref{wien}) can be written as a system of four
 matrix equations: \begin{eqnarray} \label{filter} &&{\bm W}^{11}\left({\bm
       A}^{11}{\bm\Delta}^T {\rm tr}({\bm A}^{11}) + {\bm N}^T \right)
 \nonumber\\
& + & {\bm W}^{12}{\bm A}^{22}{\bm\Delta}^{TE} {\rm tr}({\bm A}^{11}) =
 {\bm\Delta}^T{\rm tr}({\bm A}^{11}) \\
\label{filter1}
&&{\bm W}^{12}\left({\bm A}^{22}{\bm\Delta}^E {\rm tr}({\bm A}^{22}) + {\bm N}^E
\right) \nonumber\\
&+& {\bm W}^{11}{\bm A}^{11}{\bm\Delta}^{TE} {\rm tr}({\bm A}^{22}) =
 {\bm\Delta}^{TE}{\rm tr}({\bm A}^{22}) \\
\label{filter2}
&&{\bm W}^{21}\left({\bm A}^{11}{\bm\Delta}^T {\rm tr}({\bm A}^{11}) + {\bm N}^T
\right) \nonumber\\
&+& {\bm W}^{22}{\bm A}^{22}{\bm\Delta}^{TE} {\rm tr}({\bm A}^{11}) =
 {\bm\Delta}^{TE}{\rm tr}({\bm A}^{11}) \\&& {\bm W}^{22}\left({\bm
    A}^{22}{\bm\Delta}^E {\rm tr}({\bm A}^{22}) + {\bm N}^E \right)
\nonumber\\
&+& {\bm W}^{21}{\bm A}^{11}{\bm\Delta}^{TE} {\rm tr}({\bm A}^{22})  = 
{\bm\Delta}^E{\rm tr}({\bm A}^{22})
\label{filter3}
\end{eqnarray}
These equations can be solved by substitution. One can readily verify that the
equation for ${\bm W}^{11}$ reduces to the form earlier derived by Bouchet
{\em et al. \/} (1995) and Tegmark \& Efstathiou (1996) when the
cross-correlation between the fields is switched off.

Bouchet {\em et al.} (1996) defined a quantity they termed 'quality factor' to
quantify the merit of extraction of the signal corresponding to a given
process.  A straightforward generalization of this quality factor, valid for
multiple fields, can be written as:
\begin{equation}
Q_{pp'}^{ij}=\frac{\langle \hat x_p^i \hat x_{p'}^j\rangle }{\langle x_p^i
x_{p'}^j\rangle } = W_{p\nu}^{ik} A_{\nu p''}^{kl} \langle x_{p''}^l
x_{p'}^j\rangle
\label{qual}
\end{equation}
where we have used Eq.~(\ref{wien}) to write the second equality. Since this
is is just the ratio of the spectra of the minimum-variance estimated map to
the spectra of the real one, it can be viewed as the effective window function
of the experiment.

Eq.~\ref{qual} can be expanded to yield: \begin{eqnarray} \label{qual1} {\bm
 Q}^{11} & = & \left( {\bm\Delta}^T \right)^{-1} \left( \right.  \left. {\bm
 W}^{11}{\bm A}^{11}{\bm\Delta}^T + \right.  \left. {\bm W}^{12}{\bm
 A}^{22}{\bm\Delta}^{TE} \right) \\
\label{qual2}
{\bm Q}^{22} & = & \left( {\bm\Delta}^E \right)^{-1} \left( \right.
\left. {\bm W}^{21}{\bm A}^{11}{\bm\Delta}^{TE} + \right.  \left. {\bm
W}^{22}{\bm A}^{22}{\bm\Delta}^E \right) \\
\label{qual3}
{\bm Q}^{12} & = & \left( {\bm\Delta}^{TE} \right)^{-1} \left( \right.
\left. {\bm W}^{11} {\bm A}^{11}{\bm\Delta}^{TE} + \right.  \left. {\bm
W}^{12}{\bm A}^{22}{\bm\Delta}^E \right) \\
\label{qual4}
{\bm Q}^{21} & = & {\bm Q}^{12}
\end{eqnarray}
${\bm Q}^{11}$ and ${\bm Q}^{22}$ can readily be interpreted as the quality of
the reconstruction of temperature and polarisation maps, respectively.  As
expected in the presence of cross-correlations, the quality factor of either
field is better than in the case without cross-correlations. Although the
reconstruction of temperature maps is only slightly changed by the
cross-correlation term (the term proportional to ${\bm W}^{12}$ in
Eq.~(\ref{qual1})), the quality of the polarisation reconstruction gets a big
boost from the presence of temperature-polarisation cross-correlation and
almost half the contribution to ${\bm Q}^{22}$ comes from the ${\bm W}^{21}$
term in Eq.~(\ref{qual2}).  However the meaning of the term ${\bm Q}^{21}$
(and ${\bm Q}^{12}$) is not apparent.  Much of the contribution to ${\bm
Q}^{12}$ comes from the term with ${\bm W}^{11}$, and therefore it is very
close to the quality factor for the extraction of temperature and is nearly
independent of the polarisation noise.  It is not surprising as it merely
tells us that to optimally reconstruct the cross-correlation one needs to
throw out the noisy data, i.e. the polarisation information!

However, the quantity of interest is the error in the extraction of the power
spectrum of cross-correlation which should not be confused with ${\bm
Q}^{12}$.  To get a real idea of the error bars of the different spectra, we
must define estimators of those power spectra from the filtered data, and
compute their covariances. As we shall see, while ${\bm Q}^{11}$ and ${\bm
Q}^{22}$ directly give the covariance of the E and T power spectra, a more
complicated expression is needed for the covariance of the ET power spectrum.

\subsection{Unbiased estimators of power spectra}

Eq.~(\ref{recon}) is the data obtained after performing Wiener filtering on
the multi-frequency maps. Our aim in this section is to use this data to write
an unbiased estimator of the power spectra, which is defined such that
$\langle \hat C_p^{ij}\rangle = \langle x_p^i x_p^j\rangle$.  From
Eqs. (\ref{fil}) and (\ref{recon}), the power spectrum of $\hat x_p^i$ can be
written as:
\begin{equation}
\langle \hat x_p^i \hat x_p^j \rangle = W_{p\nu}^{il} W_{p\nu '}^{jm} \left
 [A_{\nu p'}^{ln} A_{\nu ' p''}^{mq} \langle x_p^n x_p^q \rangle + \langle
 b_\nu^l b_{\nu '}^m \rangle \right ],
\end{equation}
which can be expressed as:
\begin{eqnarray}
\langle \hat x_p^1 \hat x_p^1 \rangle & = & (Z_p^{11} C_p^T + b_p^{11}) =
Q_p^{11} C_p^T \nonumber \\ \langle \hat x_p^2 \hat x_p^2 \rangle & = &
(Z_p^{22} C_p^E + b_p^{22}) = Q_p^{22} C_p^E \nonumber \\ \langle \hat x_p^1
\hat x_p^2 \rangle & = & (Z_p^{12} C_p^{TE} + b_p^{12}) = Q_p^{12} C_p^{TE}.
\label{break}
\end{eqnarray}
The meaning of the first equality in 
 Eqs.~(\ref{break}) can be easily understood: the
Wiener-filtered power spectrum for a given process can be expressed in terms
of the true underlying power spectrum smeared by foregrounds ($Z_p^{ij}$
terms) plus the noise. However, unlike the case with no foregrounds,
 The 'noise' terms $b_p^{ij}$ have contribution
not only from the instrumental noise but also from residual foregrounds from
other processes and fields. The second equality comes from Eq.~(\ref{qual}).
 These equations can then be used to write unbiased
estimators of the power spectra:
\begin{eqnarray}
\hat C_p^T & = & \frac{1}{Z_p^{11}} \left( \frac{1}{2\ell +1}\sum_{m} \| \hat
x_p^1(m) \hat x_p^1(m)\| -b_p^{11} \right) \\ \hat C_p^E & = &
\frac{1}{Z_p^{22}} \left( \frac{1}{2\ell +1}\sum_{m} \| \hat x_p^2(m) \hat
x_p^2(m)\| -b_p^{22} \right) \\ \hat C_p^{TE} & = & \frac{1}{Z_p^{12}} \left(
\frac{1}{2\ell +1}\sum_{m} \| \hat x_p^1(m) \hat x_p^2(m)\| -b_p^{12} \right)
\label{unbiased}
\end{eqnarray}
From the unbiased estimators, one can readily compute the covariances of the
various power spectra:
\begin{eqnarray}
  \label{covar1} 
{\rm \bf Cov}(\hat C_p^T) & = & \frac{2}{2\ell+1}\left( C_p^T
Q_p^{11}/Z_p^{11} \right)^2 \\ 
\label{covar2} 
{\rm \bf Cov}(\hat C_p^E) & = &
\frac{2}{2\ell+1}\left( C_p^E Q_p^{22}/Z_p^{22} \right)^2 \\ {\rm \bf
Cov}(\hat C_p^{TE}) & = & \frac{1}{2\ell+1}\frac{\left(
(Q_p^{12})^2(C_p^{TE})^2 + Q_p^{11}Q_p^{22}C_p^T C_p^E \right)}{(Z_p^{12})^2}
\label{covar3}
\end{eqnarray}
Given the instrumental noise and the expected level of foregrounds,
Eqs.~(\ref{covar1}), (\ref{covar2}), and~(\ref{covar3}) can be used to
estimate the precision with which the power spectra of various quantities can
be determined. In the next section, we use the specifications of future
experiments MAP and \plancks to estimate the fractional errors on the power
spectra. In calculating the covariances, we have assumed all the
processes---CMB and foregrounds---to be Gaussian.  If a fraction of sky,
$f_{\rm sky}$, is covered then the corresponding expressions for the
covariances can be obtained by dividing the equations~(\ref{covar1}),
(\ref{covar2}), and~(\ref{covar3}) by $f_{\rm
sky}$. Throughout this paper we assume complete sky coverage, ie. $f_{\rm sky}
= 1$.

\section{Polarised foregrounds}

The most dominant Galactic polarised foregrounds are expected to be the
polarised components of synchrotron and dust. In addition, it is possible that
the free-free emission is also polarised at $10\%$ level, which could be a
further deterrent to extracting the CMB polarisation. We neglect the
possibility of polarised free-free emission
in this paper (for more details see Keating
{\em et al. \/} 1997). 

The galactic synchrotron radiation originates from the interaction of 
cosmic ray particles with the galactic magnetic field and
 is known to dominate  the galactic radio emission for 
frequencies $\le 10 \rm \, GHz$. In theory, this 
radiation can be $ \simeq 70 \, \%$ polarised, for 
the observed energy spectrum of the cosmic ray particles (Rybicki \& 
Lightman 1979).  However, 
because of the cancellation due to incoherent addition of polarised component 
 along any line of sight, Faraday depolarisation, and non-uniform magnetic
fields, the true polarised percentage is observed
 to be $\simeq 40 \%$ 
with a dependence on galactic latitude (Burn 1966; Spoestra 1984). 
The galactic emission in FIR and millimeter wavelengths is dominated
 by dust emission.
 The dust particles align themselves with the interstellar magnetic,
 and because the dust particles are not spherical the resulting FIR and 
millimeter emission is polarised (Hildebrand  \& Dragovan  1995).

{\em Polarised dust \/}: The Galactic dust is seen to be polarised at a small
level ($\la 10\%$). This is likely to be the major foreground at frequencies at
which the \plancks HFI will operate. Though there exist no observations of
polarised dust at high Galactic latitudes, it is possible to model this
emission using observations at smaller Galactic latitudes. Prunet {\em et
al. \/} (1998) modelled this distribution and computed the power spectra of
polarisation and temperature-polarisation cross-correlation. They showed that
the relevant power spectra can be fitted as:
\begin{eqnarray}
C_{E}^{\it dust}(\ell) &=& 8.9 \times 10^{-4}\, \ell^{-1.3} \, \, ({\rm \mu K}
)^2 \\ C_{B}^{\it dust}(\ell) &=& 1.0 \times10^{-3}\, \ell^{-1.4} \, \, ({\rm
\mu K})^2 \\ C_{ET}^{\it dust}(\ell) &=& 1.7 \times10^{-2}\, \ell^{-1.95} \, \,
({\rm \mu K})^2.
\label{dust}
\end{eqnarray}
These power spectra are normalized at $100 \, \rm GHz$.  They correspond to Galactic latitudes between $30^\circ$ and $45^\circ$ and are taken here as
representative of the all sky average
(for more details on this and other related
issues, see Prunet {\it et al. \/} 1998). The dust emissivity is assumed
to be proportional to $\nu^2$ with a temperature of $17.5 \, \rm K$ (Boulanger
{\em et al. \/} 1996). For comparison, we recall that the temperature
power spectrum at $100\,\rm GHz$ can be fitted by (BG98):
\begin{equation}
C_{T}^{\it dust}(\ell) = 176\, \,\ell^{-3}\, \, ({\rm \mu K})^2
\end{equation}

{\em Polarised Synchrotron emission\/}: This foreground will greatly undermine
the performance of MAP and \plancks LFI in the detection of CMB
polarisation. The existing maps of polarised component of Galactic synchrotron
shows this emission to be polarised at a level between $20$ to $60 \%$,
depending on the Galactic latitude, at radio frequencies $\le 1.4 \, \rm GHz$
(Spoestra 1984). These can be extrapolated to millimeter wavelengths using the
frequency dependence of synchrotron emission (Lubin \& Smoot 1981). However,
such a procedure is fraught with uncertainties: (1) The synchrotron spectrum
is not well known up to millimeter frequencies (for recent attempts to measure
the spectrum up to $10 \, \rm GHz$ see Platania {\em et al. \/} 1997), (2) At
low frequencies ($\le 1.4 \, \rm GHz$) the polarised emission is not optically
thin because of Faraday depolarisation.  This means that the observed emission
suffers from substantial depolarisation (and also the spatial distribution is
affected because of the rotation of the polarisation axis). The Faraday
depolarisation is proportional to $\nu^{-2}$ and becomes negligible at
millimeter wavelengths. Therefore, the degree of polarisation at these
frequencies is expected to be higher; in addition it will have a different
spatial distribution. So the existing data needs to be corrected for these
effects when an extrapolation to higher frequency is performed. It is not easy
to do so with the present data. For the lack of data, we assume the
synchrotron emission to be polarised at $44\%$ level with the same spatial
distribution ($\ell$-dependence) as the unpolarised emission. We assume both
$E$ and $B$ mode power spectra to correspond to this level of polarisation. It
is probably justified because the modelling of dust polarised emission also
shows comparable emission for these mode. Furthermore, we assume perfect
cross-correlation between $E$-mode polarisation and temperature. This is
likely to be the case because both the polarised and unpolarised emission
depend on the square of the magnetic field (its component in the plane of the
sky) and the emission at high Galactic latitudes is mostly dominated by one
structure.  We obtain:

\begin{eqnarray}
C_{E}^{\it syn}(\ell) &=& 0.2 \times C_{T}^{\it syn}(\ell) \\ C_{B}^{\it
syn}(\ell) &=& 0.2 \times C_{T}^{\it syn}(\ell) \\ C_{ET}^{\it syn}(\ell) &=&
0.44 \times C_{T}^{\it syn}(\ell)
\label{polari}
\end{eqnarray}
where
\begin{equation}
C_{T}^{\it syn}(\ell) = 4.5\, \ell^{-3}\,\, ({\rm \mu K})^2
\end{equation}
at $100\,{\rm GHz}$ (BG98).
One of the aims of taking high level of synchrotron polarised emission and
perfect cross-correlations is to consider the 'worst possible' case for the
performance of MAP and LFI. This case should be contrasted with the 
CMB case in which the
cross-correlation is $\simeq 1/3$ of the perfect cross-correlation. And 
therefore the assumed perfect cross-correlation in any foreground means that
the total signal is biased in favour of the foreground
 and would make the extraction of CMB cross-correlation more difficult.
 However, an important goal is also to ask how well
can these foregrounds be extracted using the same experiment that attempts to
measure the CMB polarisation. We address this question in a later section.
As mentioned above, the free-free emission can also be polarised at a small
level but for the assumed level of synchrotron emission it will be
sub-dominant to it at all frequencies.

\begin{figure}
  \centering\epsfig{figure=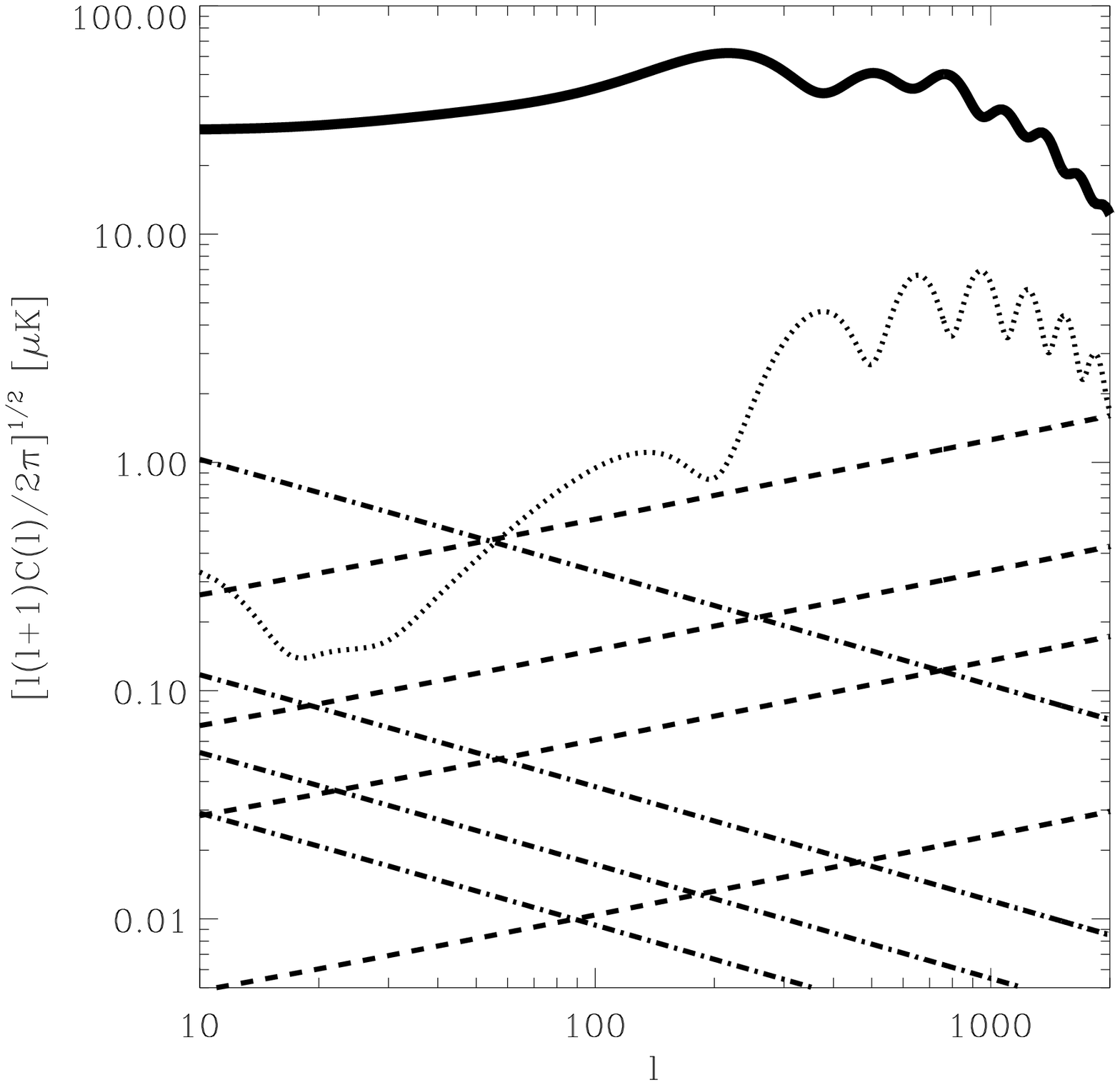,width=6cm,height=5cm}
  \centering\epsfig{figure=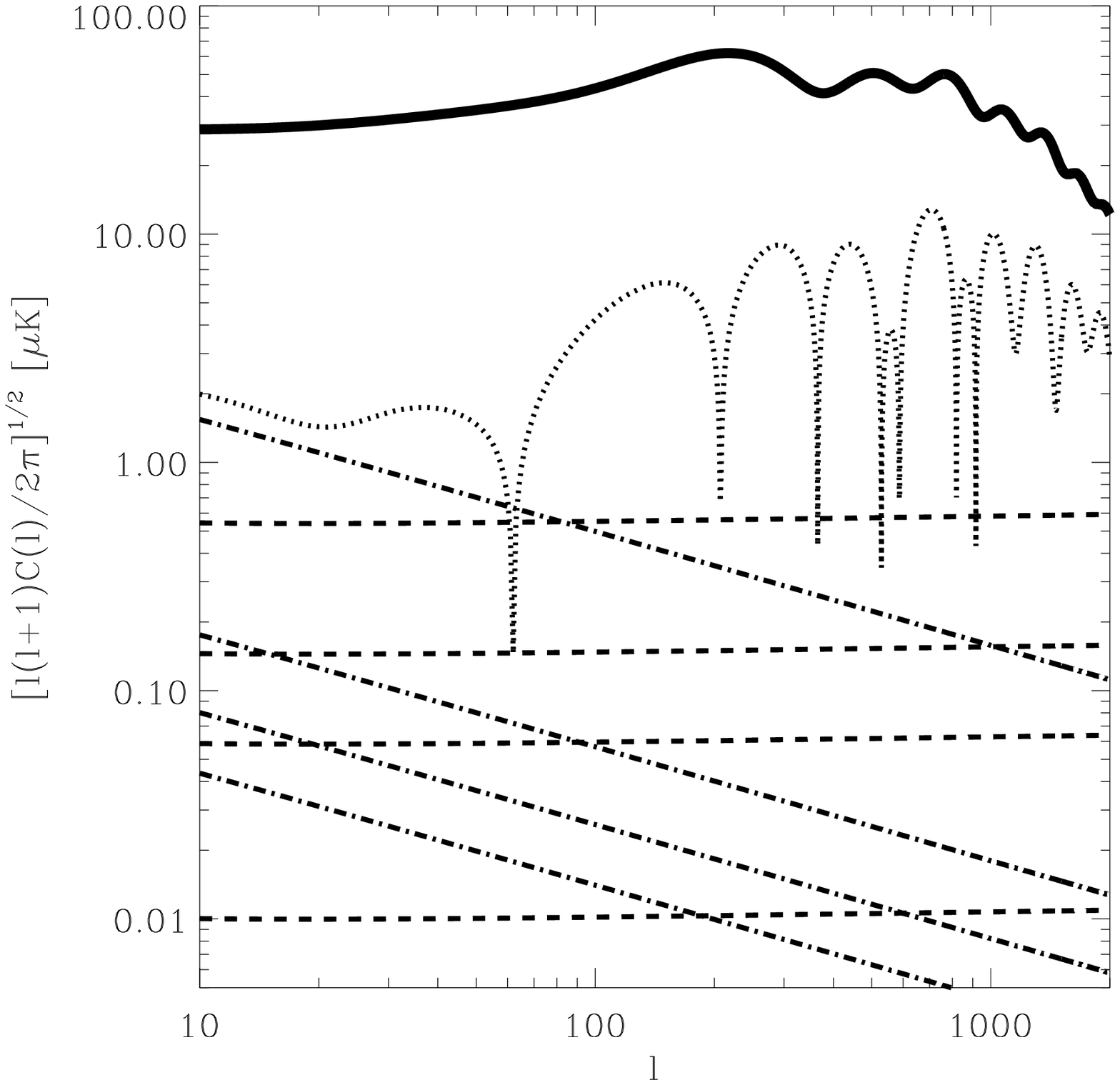,width=6cm,height=5cm}
\caption{{\em First Panel \/} : The power spectrum
  of the expected $E$-mode CMB signal is  shown along with
  the dust and synchrotron polarised foregrounds at different
  frequencies. 
  The {\em dotted \/} line  show the CMB $E$-mode power spectra. 
  The {\em dashed  \/} lines correspond to dust polarised emission
  (Eq.~(\ref{dust}) while the {\em dot-dashed \/} curves
  correspond to polarised synchrotron (Eq.~(\ref{polari})). The four
  lines for foregrounds give the expected level of contamination at
  $44 \, \rm GHz$,  $100 \, \rm GHz$,  $143 \, \rm GHz$ and  $217 \,
  \rm GHz$. The corresponding curves at these frequencies
  go from top to bottom (bottom to top) for synchrotron (dust) as the
  frequency is increased. The {\em thick solid} line gives the
  temperature power spectrum for the same underlying model, which is
  taken to be sCDM with $\tau = 0.1$. {\em Second Panel \/}: Same as
  the First Panel  for $ET$ cross-correlations.}
\label{fore_e}
\end{figure}

In Fig.~\ref{fore_e}, we compare the expected CMB
$E$ and $ET$ signal with the assumed level of foregrounds. 
The  CMB signal  dominates the foregrounds for frequency channels between
$44 \, \rm GHz$ and $217 \, \rm GHz$. The lower frequency channels of
MAP ($22$ and $30 \, \rm GHz$)  and LFI ($30 \, \rm GHz$)
and the  higher frequency channels of HFI ($545 \, \rm GHz$) will be
dominated by polarised synchrotron and dust, respectively;
these channels will help   in  an accurate determination  of these
foregrounds.

Recently, it was argued by Draine and Lazarian (1998) that a part of
galactic foreground between $10$ and $100 \, \rm GHz$
could be contributed by non-thermal emission from rotating dust
grains. If this emission is polarized at the same level as we assume
in this paper (Eq.~\ref{dust}) then it could add to the polarised
galactic emission. However, this emission is sub-dominant to the
polarised synchrotron emission we assume in this paper. And therefore
we neglect the effect of spinning dust particles in our discussion in
this paper. 

Apart from the polarised foreground and in addition to the Galactic
unpolarised foregrounds---dust, free-free, and synchrotron--- we include
several other extragalactic unpolarised foregrounds in our study: The thermal
Sunayev-Zeldovitch effect from clusters of galaxies, the infra-red point
sources, and the radio point sources (for details see Bouchet \& Gispert
1998a).  We do not consider any extragalactic polarised foregrounds.

\section{Extraction of polarised signal}

\begin{figure}
\centering\epsfig{figure=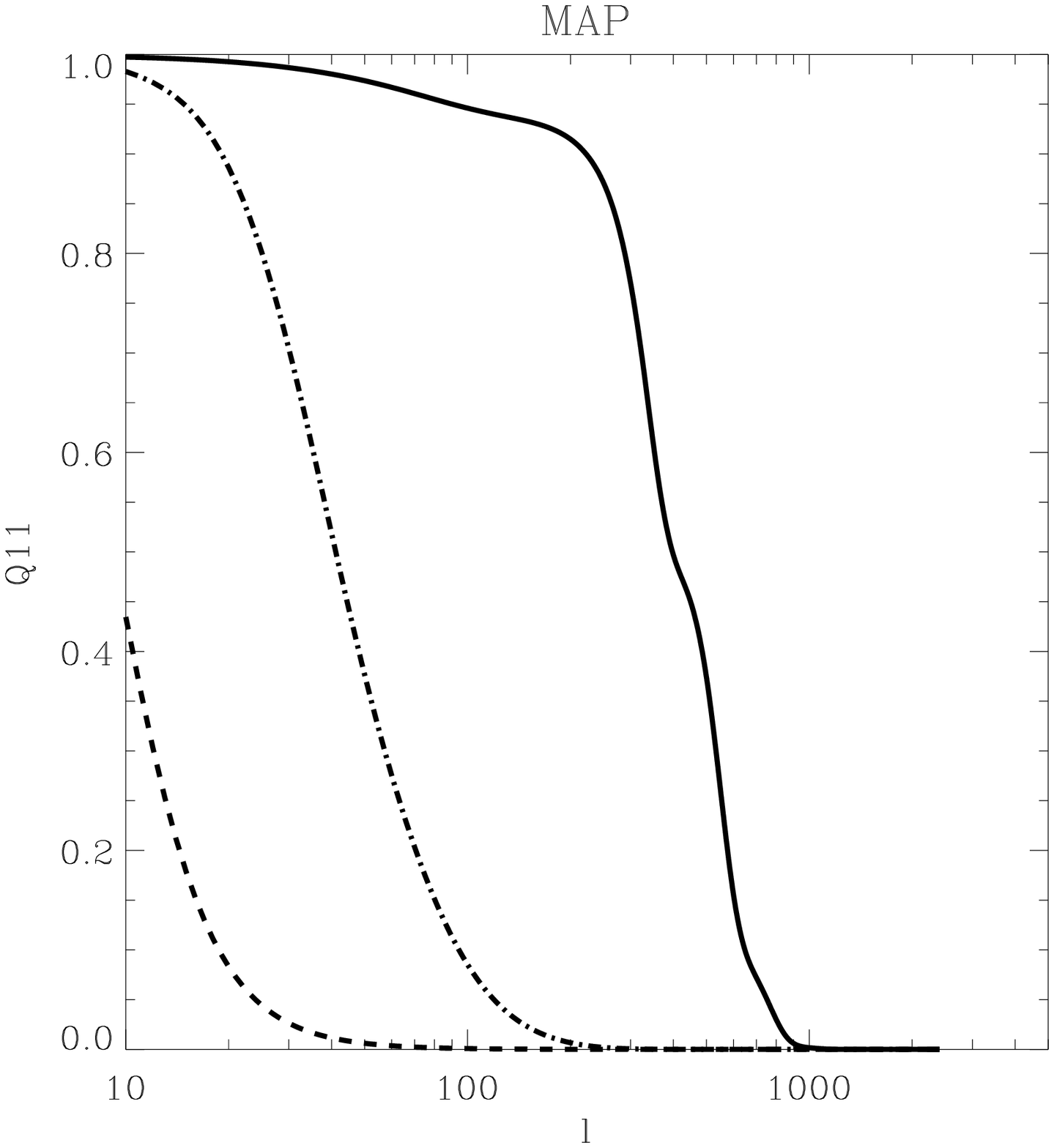,width=6cm,height=5cm}
\centering\epsfig{figure=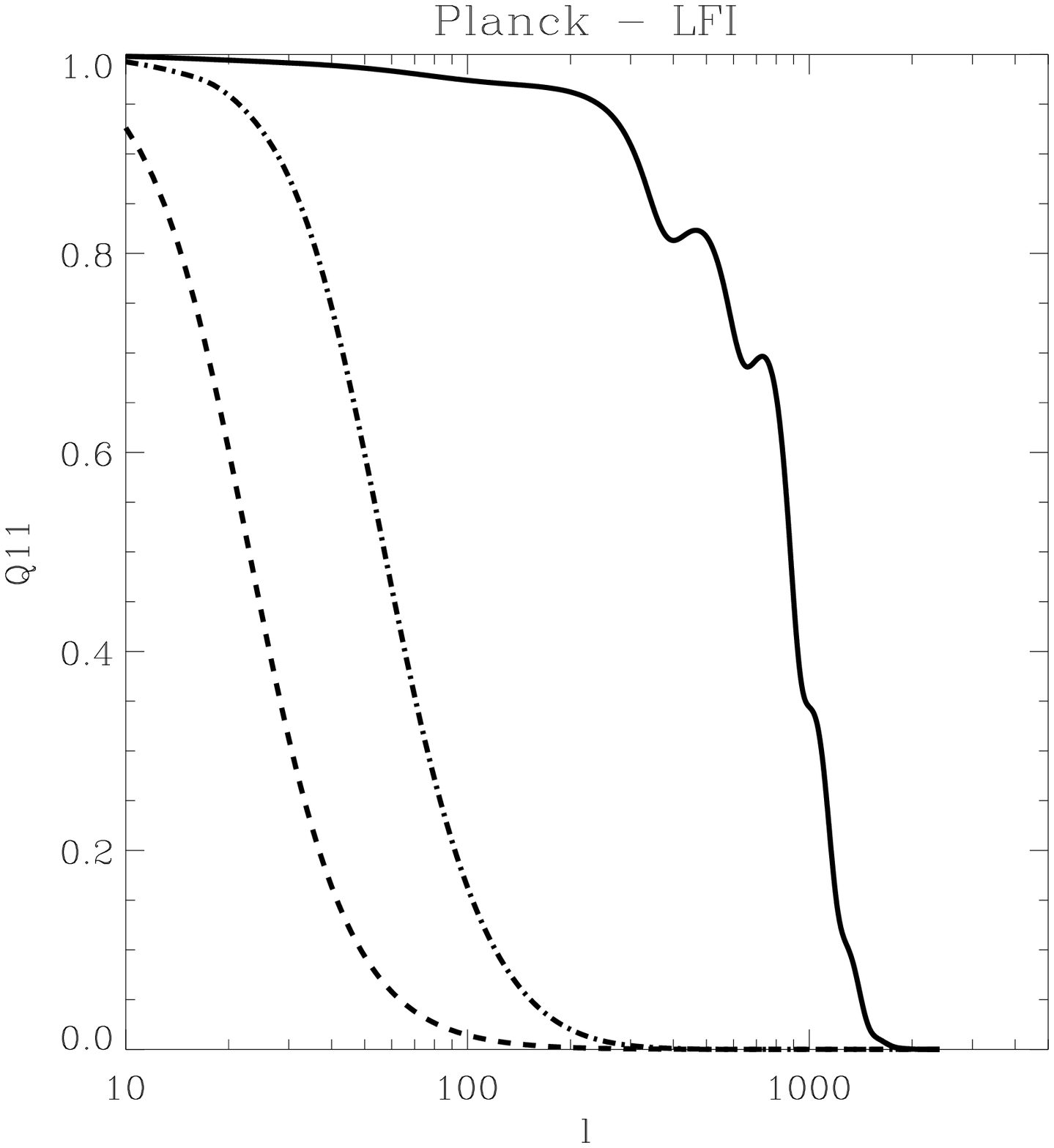,width=6cm,height=5cm}
\centering\epsfig{figure=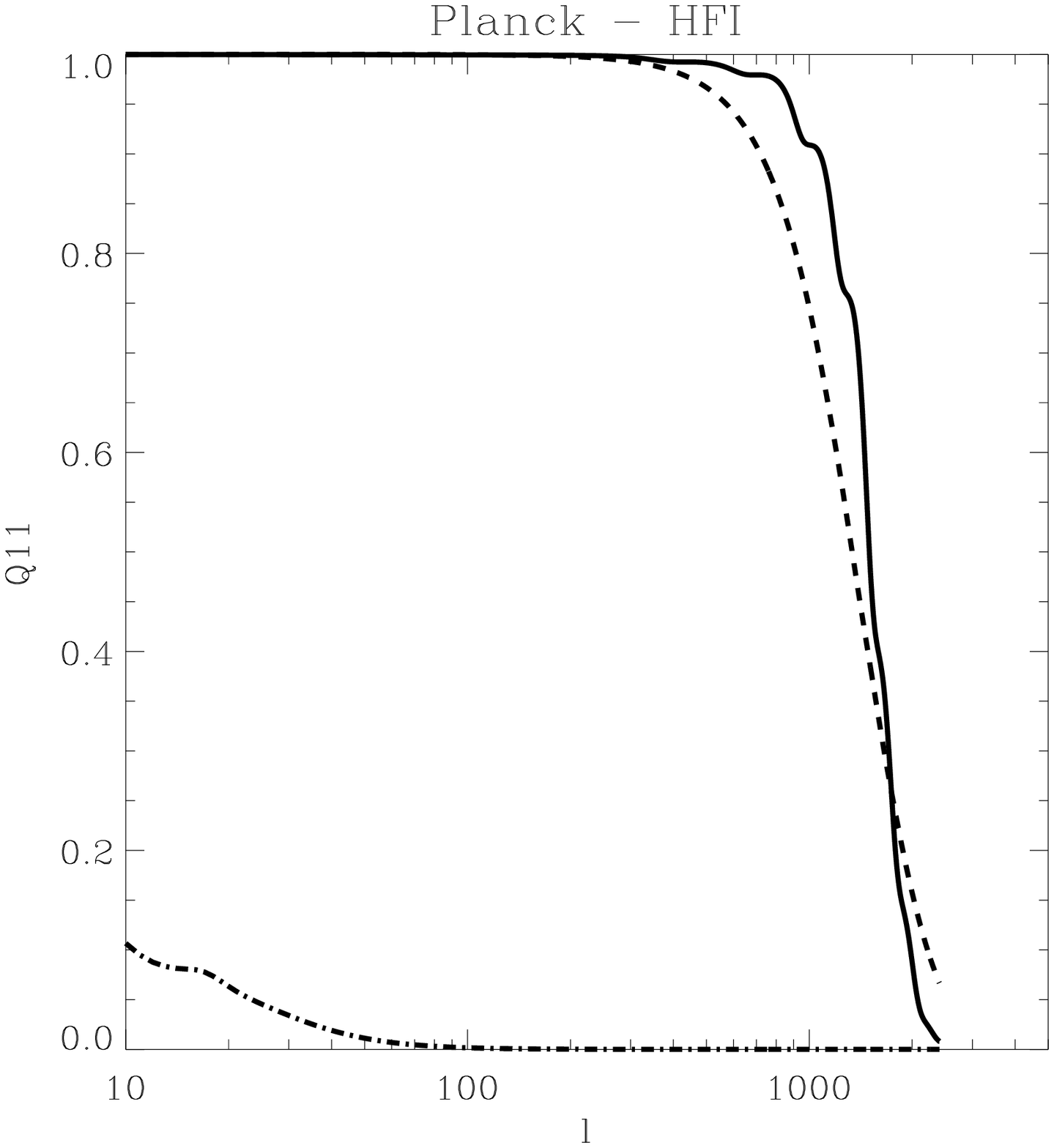,width=6cm,height=5cm}
\centering\epsfig{figure=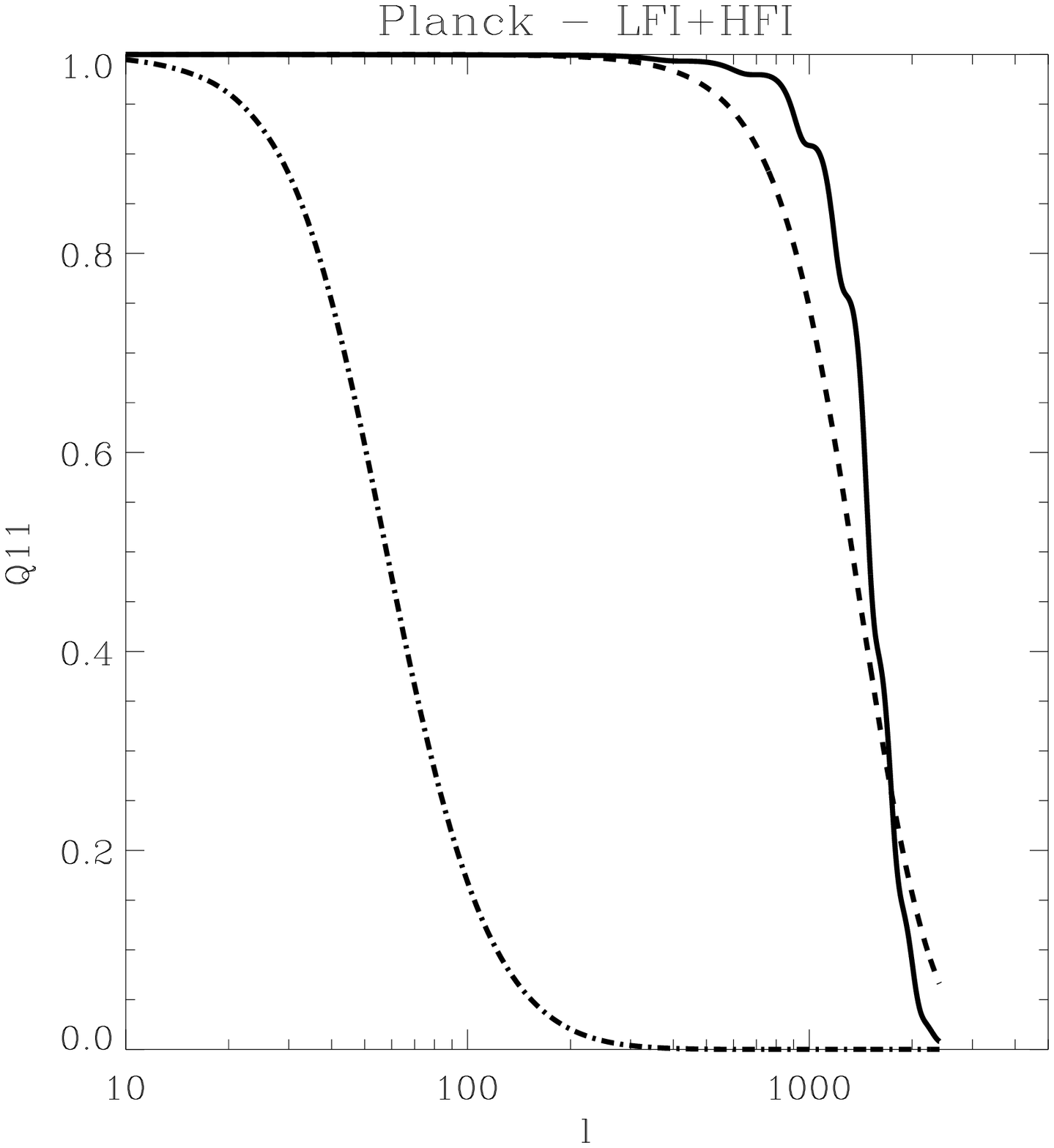,width=6cm,height=5cm}
\caption{Quality factors for the temperature fluctuations. {\em Solid}, 
  {\em dashed}, and {\em dot-dashed \/} lines stand for CMB, dust and
synchrotron respectively. As can be seen by comparing the effective windows in
the case of different experiments, despite its lower sensitivity MAP does
as well as  the LFI at recovering the synchrotron emission because of
its  larger
frequency coverage; but since this emission is relatively weak, the LFI CMB
transmission is better. Also note the improvement on the foreground
recovery when the LFI and HFI are combined, although the CMB recovery is not
improved by adding the LFI data.}
\label{qual_temp}
\end{figure}

\begin{figure}
\centering\epsfig{figure=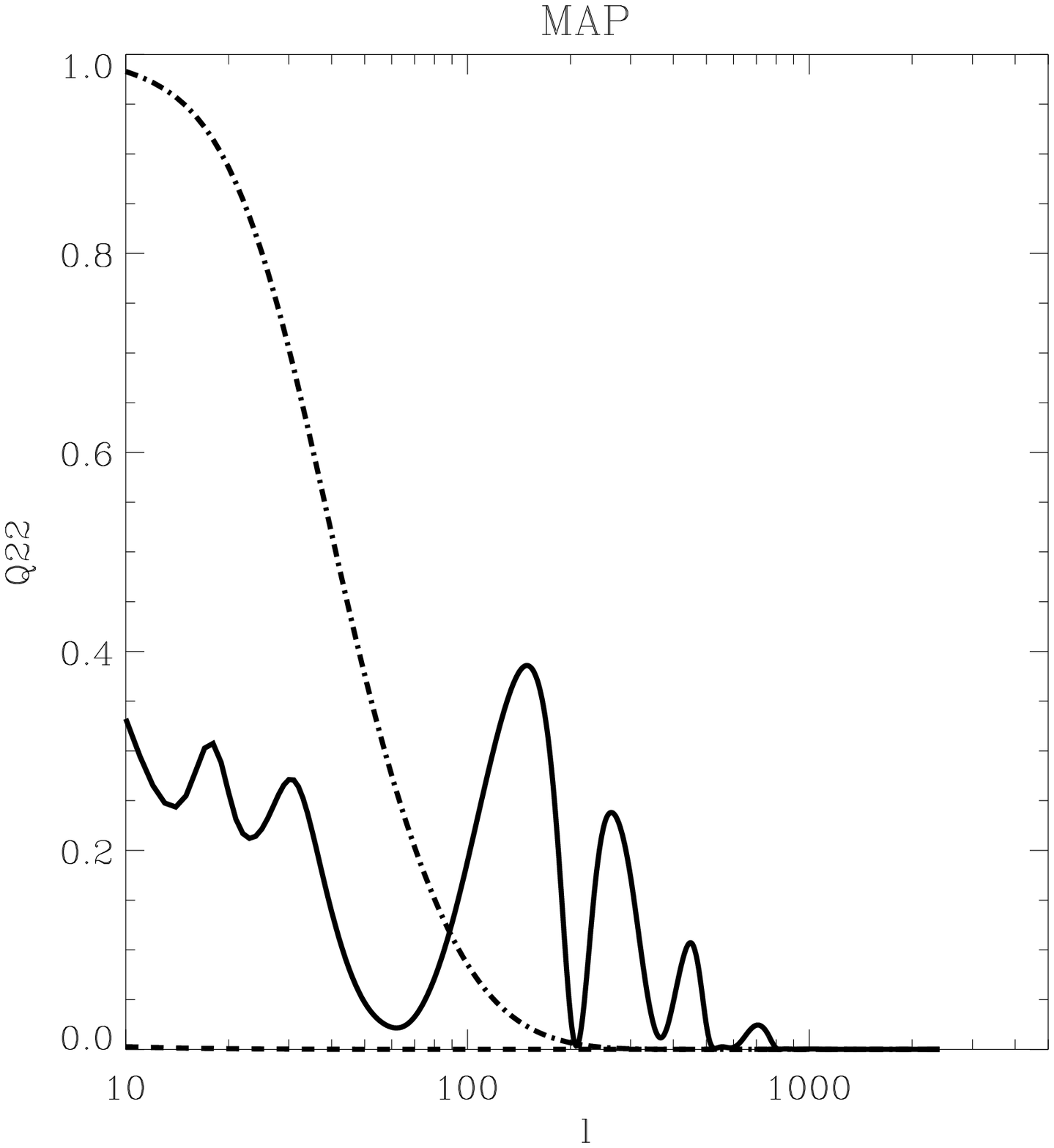,width=6cm,height=5cm}
\centering\epsfig{figure=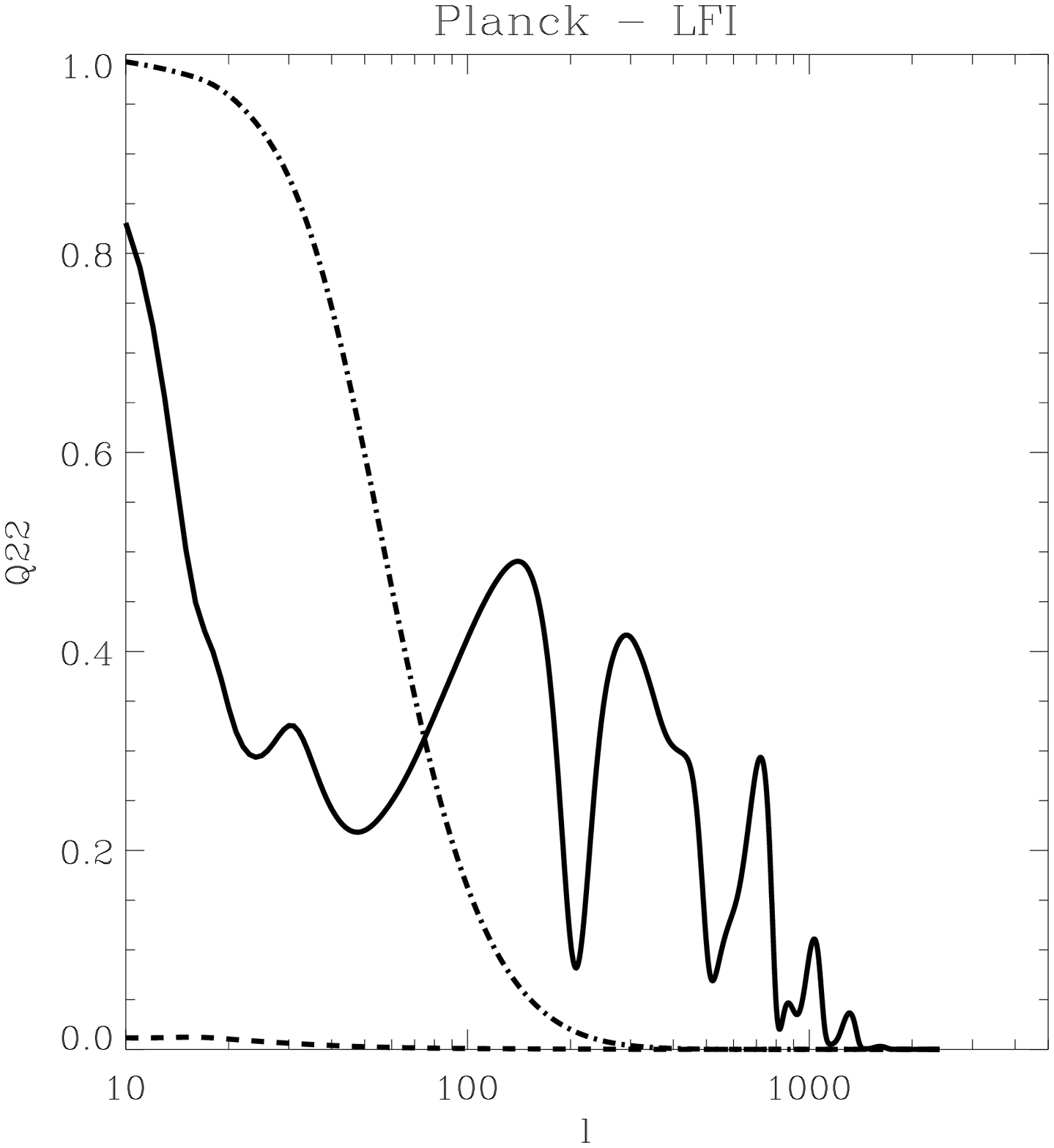,width=6cm,height=5cm}
\centering\epsfig{figure=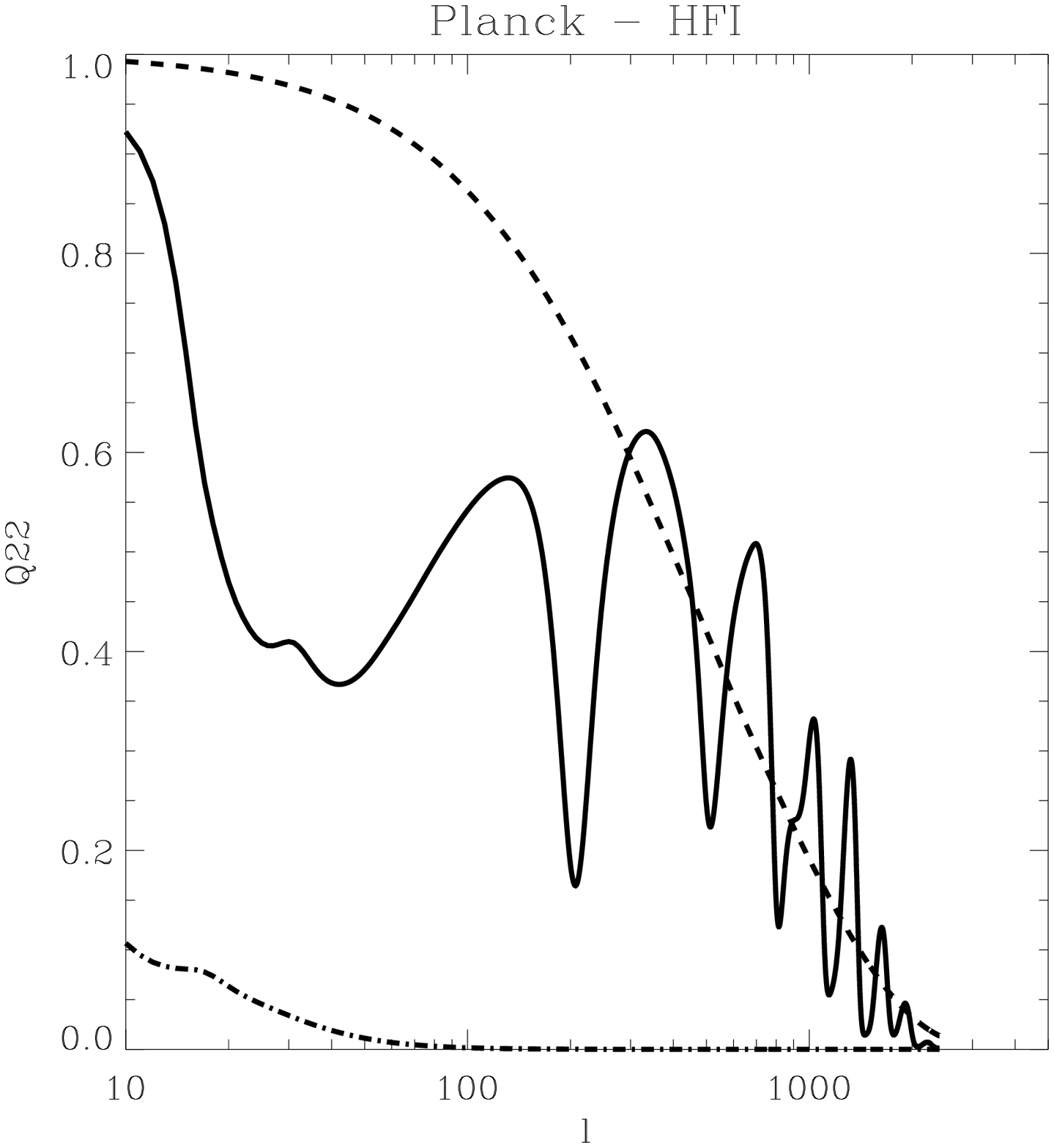,width=6cm,height=5cm}
\centering\epsfig{figure=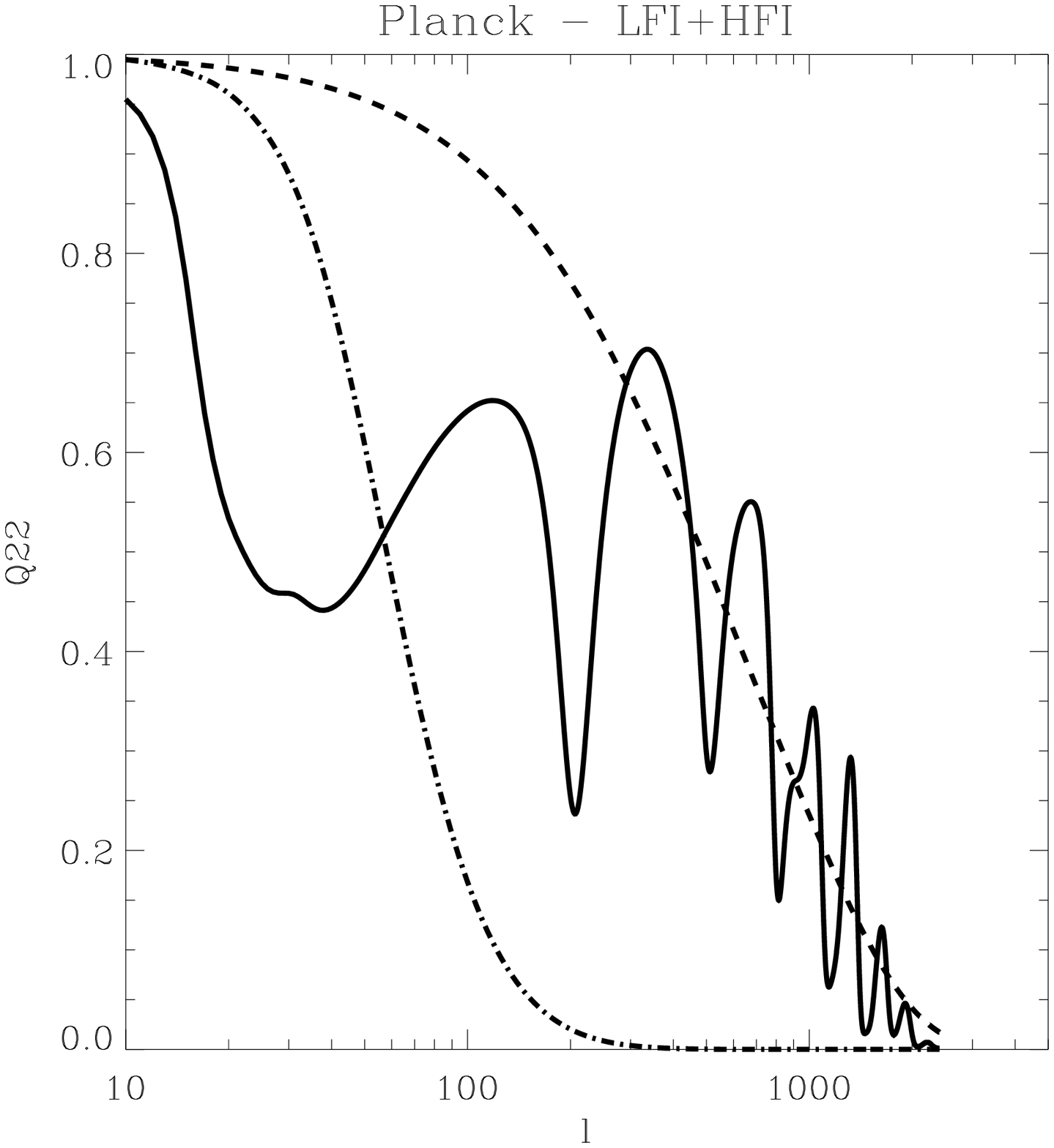,width=6cm,height=5cm}
\caption{Quality factors for the $E$-mode polarisation. {\em Solid}, {\em
dashed}, and {\em dot-dashed \/} lines stand for CMB, dust and
synchrotron respectively. Note that all the experiments extract the
foregrounds better than the CMB for small $\ell$. \plancks HFI extracts
the polarised dust better than the CMB at almost all $\ell$. This
suggest that polarised  foregrounds are likely to be extracted as well
or better than the CMB. Though
this is partly due of our assumed level of foregrounds, this feature
reflects the frequency coverage of these experiments}
\label{qual_pol}
\end{figure}

{\em Quality factors}: As mentioned in the last section, a relevant quantity
for determining the merit of signal extraction is the 'quality factor'
(Eq.~(\ref{qual})). The quality factors for temperature and $E$-mode
polarisation are shown in Figs.~(\ref{qual_temp}) and (\ref{qual_pol}) for
various processes. It should be noted that in these figures we only show the
quality factor for unpolarised synchrotron and dust though we include several
other unpolarised Galactic and extragalactic foregrounds in our analysis (see
\S 3). Throughout this and the next subsection, we take the underlying models
for calculating various power spectra to be a variant of sCDM model with $n_s
= 1$, the ratio of scalar to tensor quadruple $T/S = 0$ (no tensor signal),
and the optical depth the last scattering surface $\tau = 0.1$. Finite
optical depth to the last scattering surface enhances the polarisation signal
for $\ell \la 10$, a part of the spectrum which is of great interest
for breaking degeneracies between
various cosmological parameters (for details see Zaldarriaga, Spergel, \&
Seljak 1997). Throughout this paper,
we use the  CMB Boltzmann code  CMBFAST to generate CMB fluctuations
(Seljak \& Zaldarriaga 1996). 

As expected, the temperature signal can be extracted far more cleanly than the
polarised signal. The quality factor for the extraction of CMB drops
exponentially as the $\ell$ approaches the effective beam width of the
respective experiment. Another noticeable feature in temperature quality
factors for various experiments is that the spatial distribution of Galactic
dust emission can be discerned almost as well as the CMB signal using \plancks
HFI. This is largely  owing  to the presence of polarised channel at
 $545 \, \rm GHz$. The signal at these channels will be dominated by Galactic dust
emission; and they have sufficiently low noise levels and high enough angular
resolution to allow a good determination of the power spectra of the dust
emission up to $\ell \simeq 1000$. This should be contrasted with the
extraction of unpolarised synchrotron emission, which is one of the dominant
contaminant for $\nu \le 90 \, \rm GHz$. Unfortunately
neither MAP nor LFI can extract this
signal well. It is because (a) it does not dominate the signal at any
frequency of either MAP or LFI (b) the signal is the strongest is the lowest
resolution channels, and (c) the unpolarised synchrotron and
Galactic free-free emission have very similar spatial distribution (assumed to
be the same in this paper) and nearly the same frequency dependence, which
makes it difficult to extract either of them.

The oscillating nature of quality factors for CMB polarisation attests to the
fact that the signal is barely above the noise level.  Another important thing
to notice in the figures is the quality of extraction of processes like the
polarised component of dust and synchrotron. For HFI, the polarised dust can
be extracted better than the CMB for much of the $\ell$-range. As in the
temperature case, it is possible because of the presence of a polarised
channel at $545 \, \rm GHz$.  Equivalently, the lowest frequency channels of
MAP and LFI serve as templates for polarised synchrotron. It should be noted
that the synchrotron polarised component can be extracted much better than its
unpolarised counterpart, because, whereas the former dominates all other
signals at the lowest frequency channels of MAP and LFI, the latter, as
discussed above, doesn't dominate the total signal at any frequency, at least
in our modeling.

We shall see below how the information on quality factor translates into
errors on power spectra for various processes.

\subsection{Errors on power spectra}

\begin{figure}
\centering\epsfig{figure=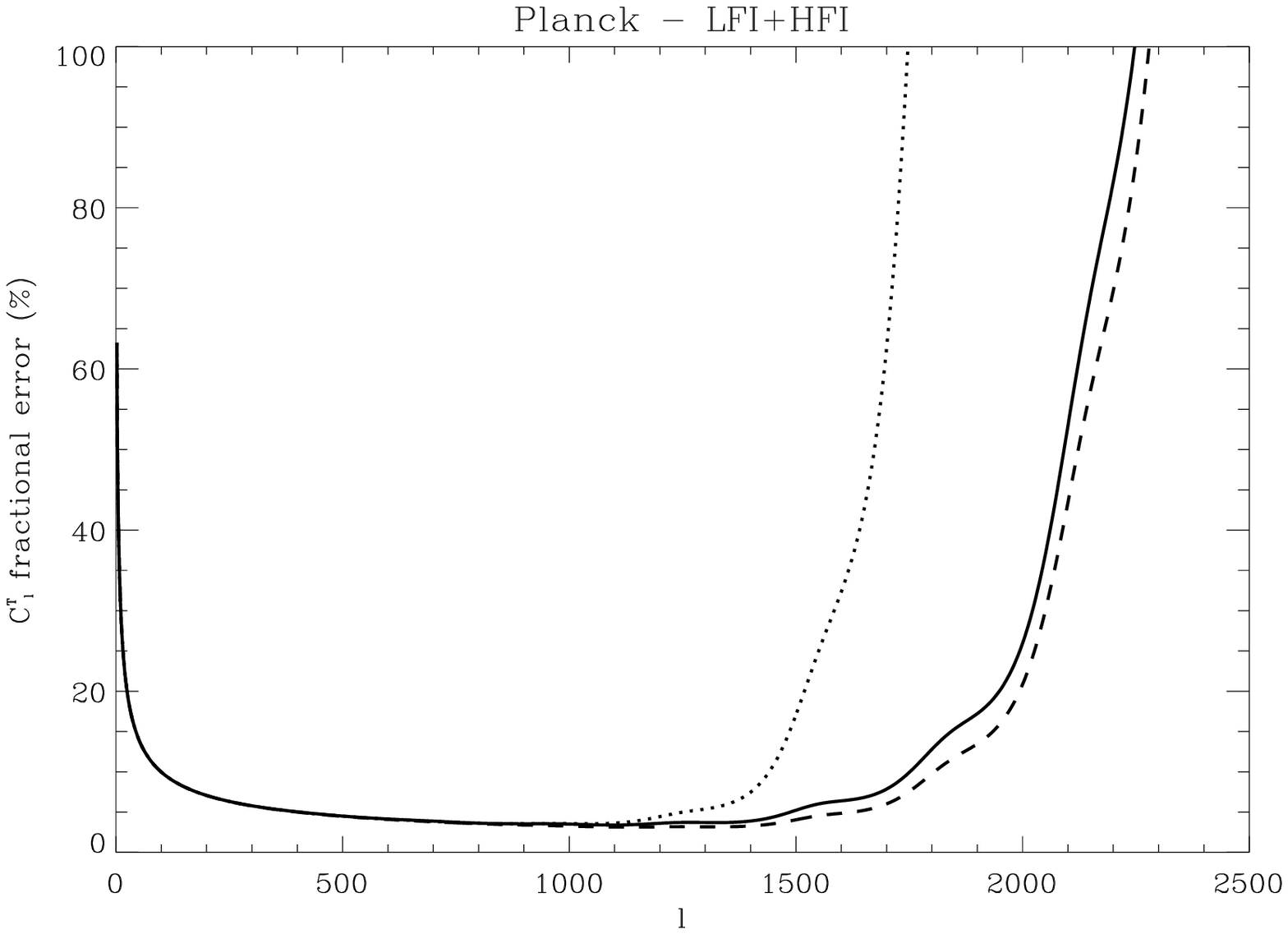,width=6cm,height=5cm}
\centering\epsfig{figure=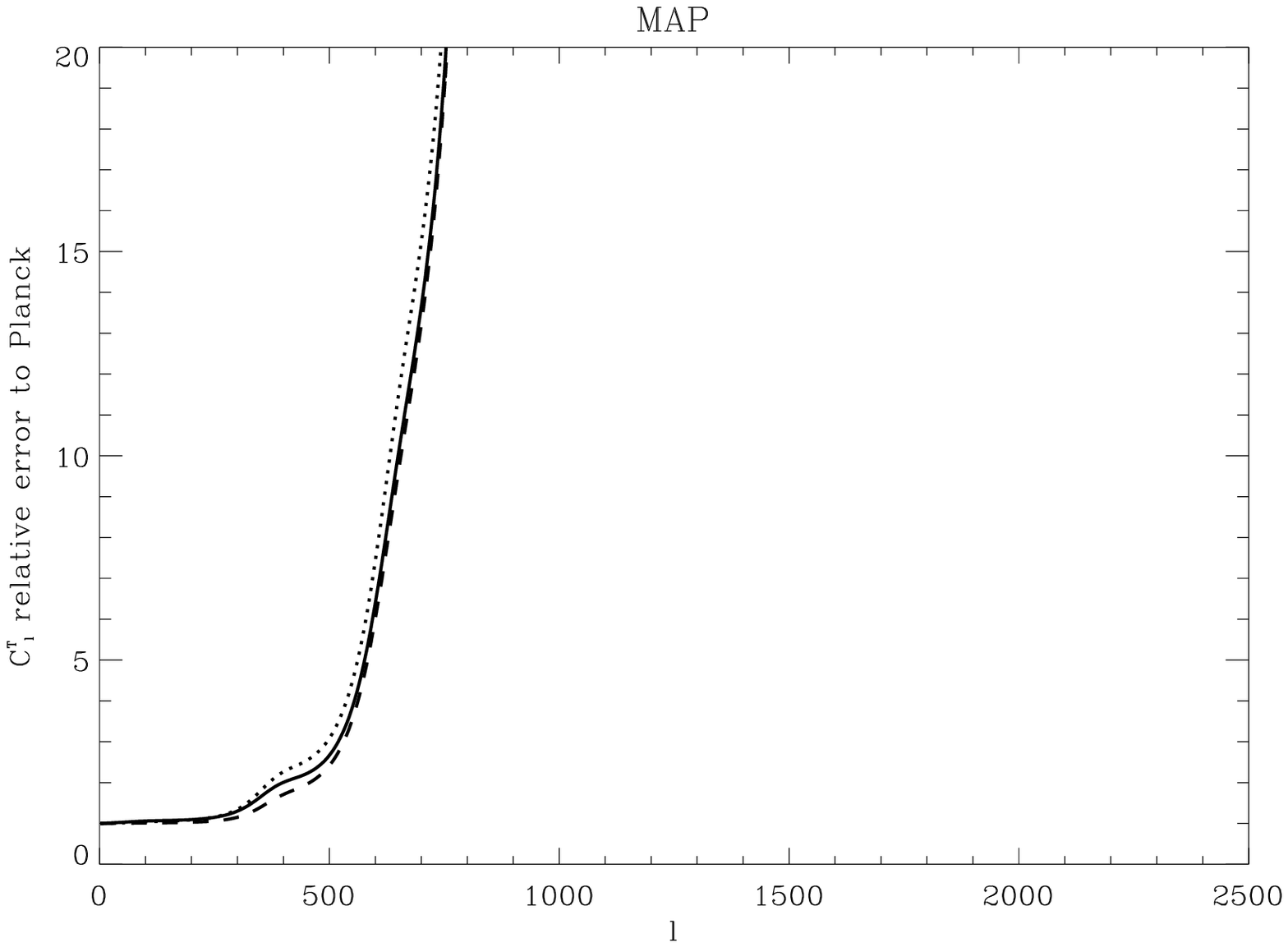,width=6cm,height=5cm}
\centering\epsfig{figure=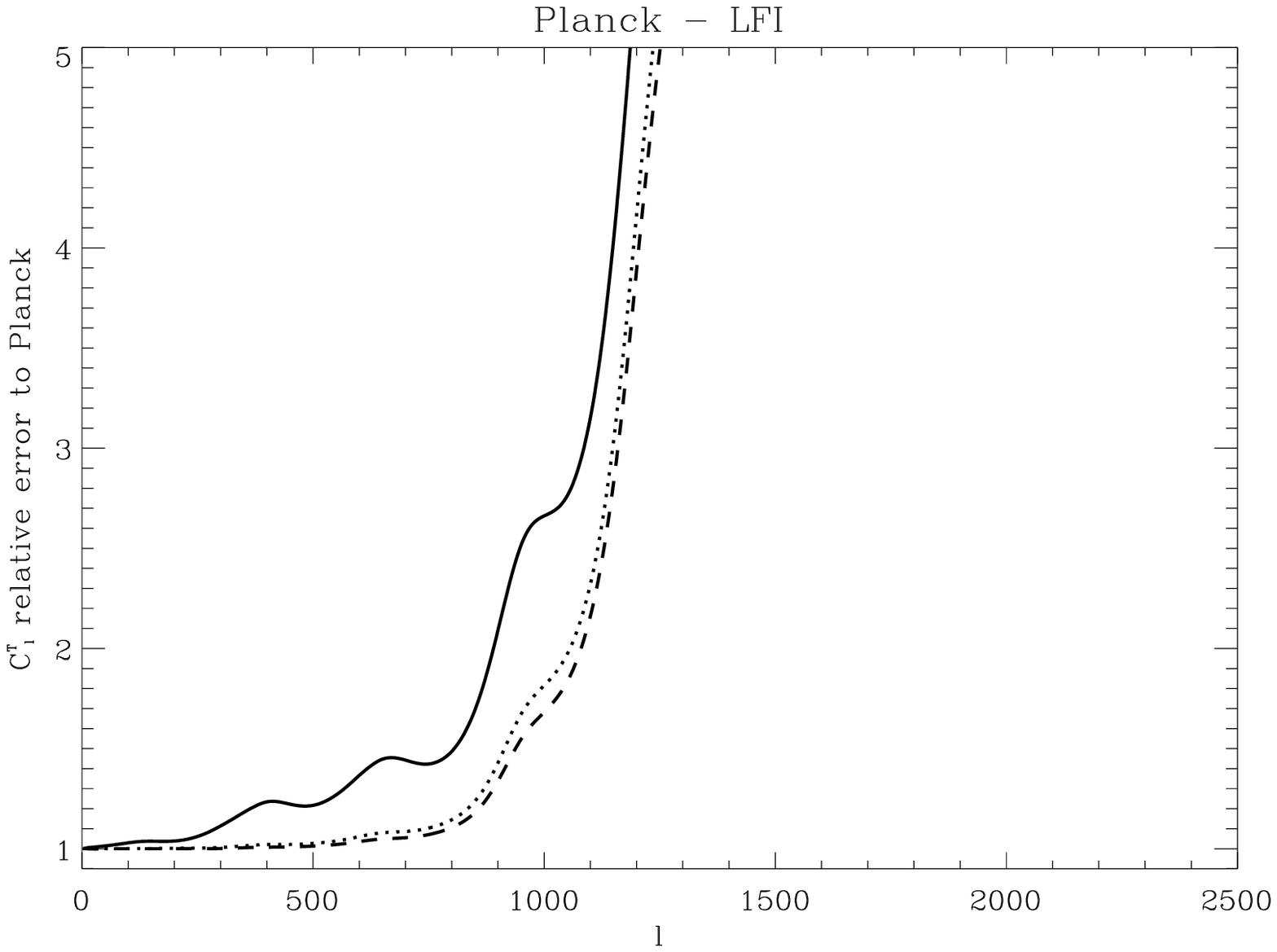,width=6cm,height=5cm}
\centering\epsfig{figure=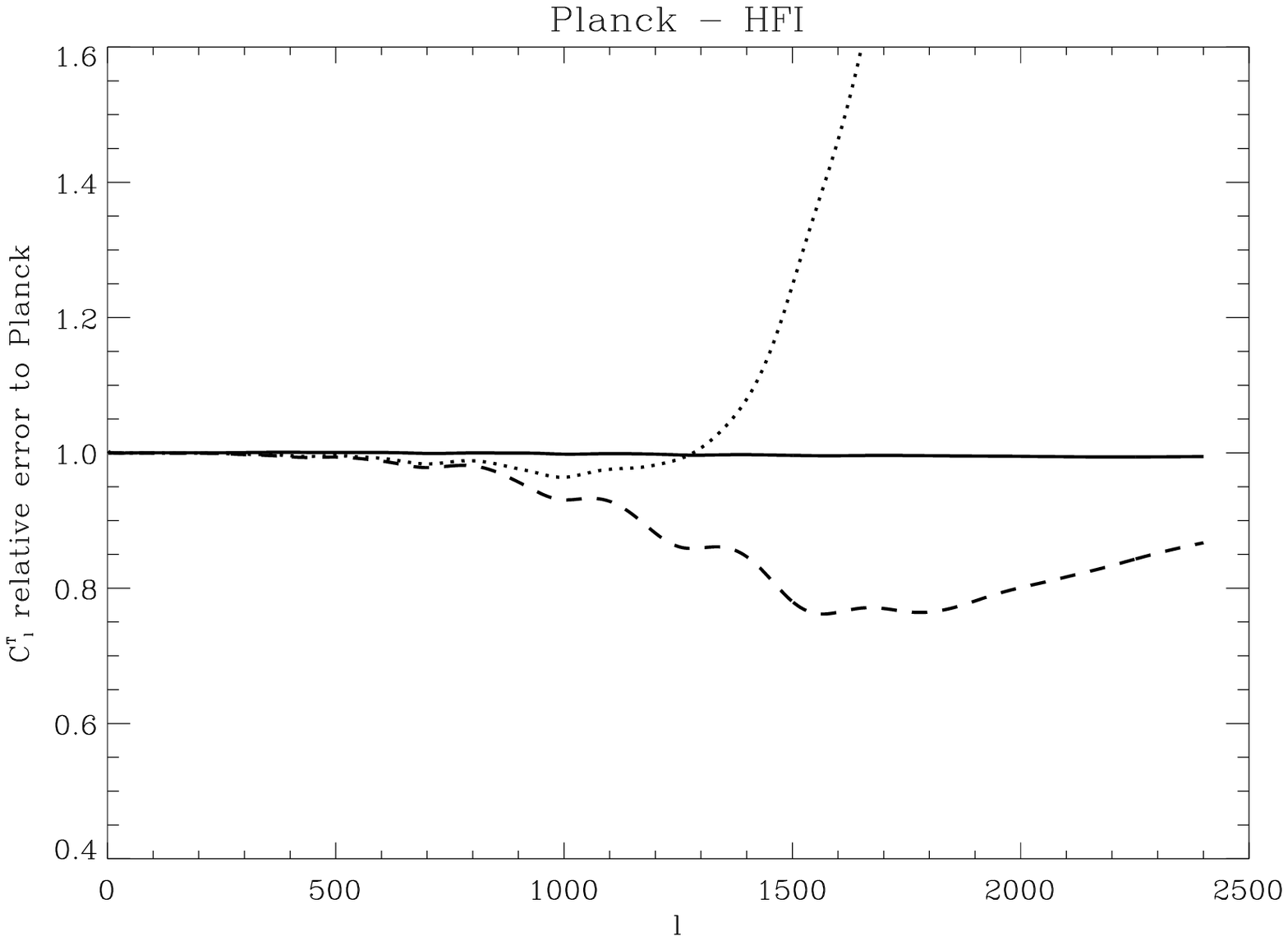,width=6cm,height=5cm}
\caption{Relative fractional error on temperature power spectra.  The first
  panel shows the expected performance of the full
\planck ~ mission in extracting this signal. The {\em Solid \/}line illustrates the
Wiener filtering case, {\em i.e. \/} the foregrounds are included and
subtracted using the Wiener filtering method described in the text.The
{\em dotted \/} line corresponds to the case if the signal is extracted
using only the best channel of \plancks ($143 \, \rm GHz$) neglecting all
the foregrounds. The other panels show the performance of other experiments
relative to the \planck -Wiener case (the {\em solid \/} line in the first
panel). In addition to the Wiener and best channel case (the same line style
as the first panel), we also show the errors using the combined sensitivity
of all the channels ({\em dashed \/} line) for various experiments in
the last three panels}
\label{cov_temp}
\end{figure}

We use Eqs.~(\ref{covar1}), (\ref{covar2}), and~(\ref{covar3}) to estimate the
fractional errors in the extraction of the CMB $E$-mode and $TE$
cross-correlation power spectra. The results are shown in
Figs.(\ref{cov_temp}), (\ref{cov_pol}), and (\ref{cov_cross}) for the
specification of various experiments. For comparison, we also plot the
expected errors using the best channel ($90 \, \rm GHz$ channel for MAP, $100
\, \rm GHz$ channel for LFI, and $143 \, \rm GHz$ channel for HFI and full
\planck) and the combined sensitivity of all the channels of each
experiment (Bond {\em et al. \/} 1997).

For temperature signal, all the experiments give similar results for $\ell \le
300$. It is expected as the signal of temperature fluctuation is so much above
the noise level for all the experiments in this $\ell$-range that additional
sensitivity does not lead to additional precision in power spectrum
estimation. The only source of error in this $\ell$-range is cosmic variance
which is obviously independent of the experiment. This information could also
be gleaned from the quality factor plots. For $\ell \ge 300$, the extraction
of temperature signal depends on the relative beam widths of relevant channels
of various experiments. As expected, HFI performs best because it will have
channels with $5'$ resolution; it is followed in performance, in that order,
by LFI and MAP. It should be noted that the results using Wiener filtering
generally lie between the performances of best channel and combined
sensitivity.

Unlike the temperature fluctuations, the extraction of polarisation and
temperature-polarisation cross-correlation depends sensitively on the pixel
noise, as is evident from comparison between fractional errors on these
quantities for various experiments. (The sharp spikes in $ET$ plots is not a
property of the extraction error but merely indicate that the signal vanishes
at these values of $\ell$ in the underlying theoretical model.)
The performances of LFI and HFI are similar for
$\ell \la 600$ largely because of comparable polarised sensitivity of the $100
\, \rm GHz$ LFI and $143 \, \rm GHz$ HFI channels. For larger $\ell$, HFI
performs better because of its higher angular resolution. Both these
experiments should extract both the E-mode power spectrum and $ET$
cross-correlation to $20 \hbox{--} 30\%$ precision, i.e. signal-to-noise,
$(C_\ell/\sqrt{{\rm \bm Cov}(C_\ell}))$, of $4\hbox{--}5$, for
$50 \la \ell \la 
1000$. The signal-to-noise is much smaller at larger scales, apart
from cosmic variance, because
the CMB signal becomes  comparable to both  foregrounds and
pixel noise. A comparison with the results from the combined
sensitivity
of all channel shows that
the presence of foregrounds degrade the signal extraction
by a factor $\le 2$ for much  of the $\ell$ range. 

From the figures, it is clear that MAP is hampered not only by foregrounds but
also by its sensitivity. If the foregrounds were neglected the combined
sensitivity of all its channels might enable a marginal detection of this
signal. However, our analysis shows that the presence of foregrounds will
degrade this detection by a factor $\simeq 4$,
and the resultant fractional errors on $E$ and
$ET$ cross-correlation power spectra will be $\ge 300 \%$ in most of
$\ell$-range except $\ell \le 10$.  And therefore it seems unlikely that MAP
could give a positive detection of either $E$ and $ET$ signal for {\em
individual} 
modes.

\begin{figure}
\centering\resizebox{\hsize}{!}{\epsfig{figure=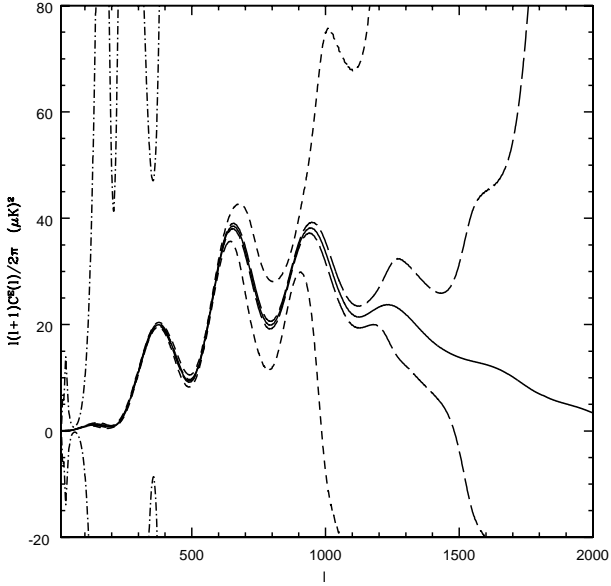}}
\caption{$E$-mode band power estimates. We plot the
  $E$-mode signal for the underlying model ({\em solid \/} line)
  and show the expected
  $1\sigma$ measurements ({\em dashed }, {\em dotted} and {\em
    dot-dashed} line for HFI, LFI and MAP respectively)
  when band power with a logarithmic interval $\Delta
  \ell /\ell = 0.2$  is taken.}
\label{band_e}
\end{figure}

\begin{figure}
\centering\resizebox{\hsize}{!}{\epsfig{figure=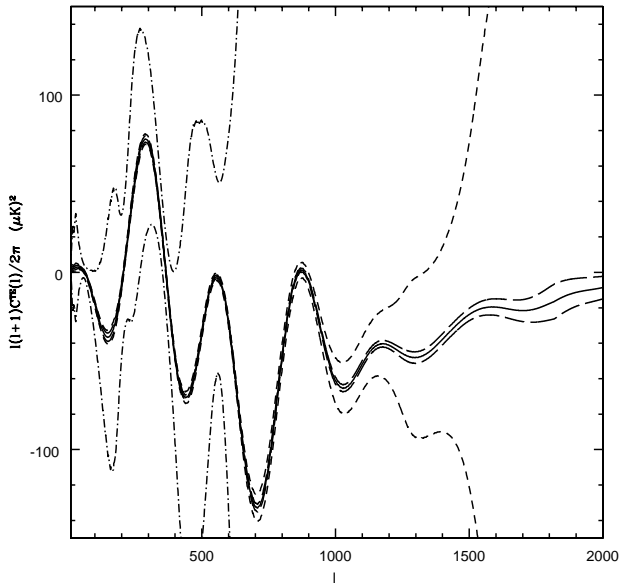}}
\caption{Same as Fig.~\ref{band_e} for the $ET$
  cross-correlation.}
\label{band_te}
\end{figure}

{\em Band-Power estimates  \/}: A way to  reduce the 'noise' further  is by
considering band-power estimates around a given $\ell$ (see Bond 1996 for
relevant definitions).  Band power taken over a logarithmic interval
of $\Delta \ell / \ell$  results in the reduction of errors by a
factor of
$\sqrt \Delta \ell$ around any $\ell$.
In Figs~(\ref{band_e}) and (\ref{band_te}), 
expected $1\sigma$ measurements  are plotted for various experiments,
with band powers taken  over a logarithmic interval of $20\%$, i.e.
$\Delta \ell / \ell =
0.2$ for \plancks HFI and LFI. As is clearly seen in the pictorial
representation of these figures, LFI and HFI should make a fairly accurate
detection   of $E$ and $TE$ power spectra for $\ell \la 750$.
For MAP these estimates show that
the errors for $ET$ signal drop to $\la 100 \%$
near the Doppler peak at $\ell \simeq  300$.  

It should be borne in mind that in the absence of any data on the power
spectrum of polarised synchrotron we took the synchrotron power spectrum to
have the same shape as the temperature power spectrum with $44 \%$
polarisation and a perfect cross correlation between $E$ and $T$. By reducing
the level of synchrotron foreground and the $ET$ cross-correlation it should
be possible to get lower errors. On the other hand, a higher level of
polarised synchrotron should allow a better extraction of synchrotron itself,
which might be subtracted using methods other than Wiener
filtering. Therefore, there is trade-off between a small and large assumed
level of foregrounds: a sub-dominant foreground will give smaller formal
errors on the extraction of CMB polarisation but these foregrounds themselves
will be elusive, thereby making it harder to quantify the errors; on the other
hand, a larger foreground can be detected and probably subtracted more
efficiently using, for instance, its non-Gaussian nature.  These
considerations makes it worthwhile to quantify the errors on foreground
extraction.

\subsection{Extraction of polarised foregrounds}

One of the important goals of future multi-frequency experiments is to {\em
detect \/} and subtract the foregrounds which hinder the determination of CMB
signal. At present one has to extrapolate spectral information on various
foregrounds from radio and FIR wavelengths to millimeter and sub-millimeter
wavelengths at which the CMB experiments operate.  A similar extrapolation is
required on the spatial distribution of foregrounds. For instance, the
synchrotron spectrum is determined only for $\nu \la 10 \, \rm GHz$
(Platania {\em et al. \/} 1997) while the spatial information is known for
angular scales $\ge 0.5^{\circ}$ from the existing all-sky maps (Haslam {\em
et al. \/} 1981). While dust maps are available with
an angular resolution of $5'$ one needs to extrapolate the dust spectrum from
$60 \, \mu m$ to millimeter wavelengths for a comparison with the CMB signal
(Neugebauer {\em et al. \/} 1984). It has been pointed out that such an
extrapolation, though useful, might result in large errors in a
high-sensitivity CMB experiment (Brandt {\em et al. \/} 1994).  Therefore, it
is of paramount importance to determine the spectrum and spatial distribution
of the foregrounds from the same experiment which attempts to measure the CMB
anisotropies.

\begin{figure}
\centering\epsfig{figure=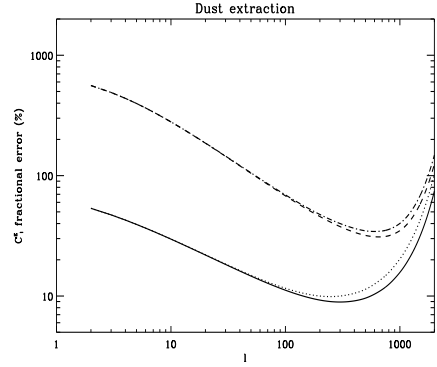,width=6cm,height=5cm}

\caption{Fractional errors on the extraction of dust $E$ and $ET$ power
spectra with HFI and full \plancks. The lower two curves correspond to $E$
power spectrum ({\em solid \/} and {\em dotted \/} lines are for HFI and
\plancks respectively) while the upper curves give the errors for $ET$
cross-correlation, with the {\em dashed \/} line for \plancks and the {\em
dot-dashed \/} line for \plancks HFI.}
\label{cov_dust}
\end{figure}

\begin{figure}
\centering\epsfig{figure=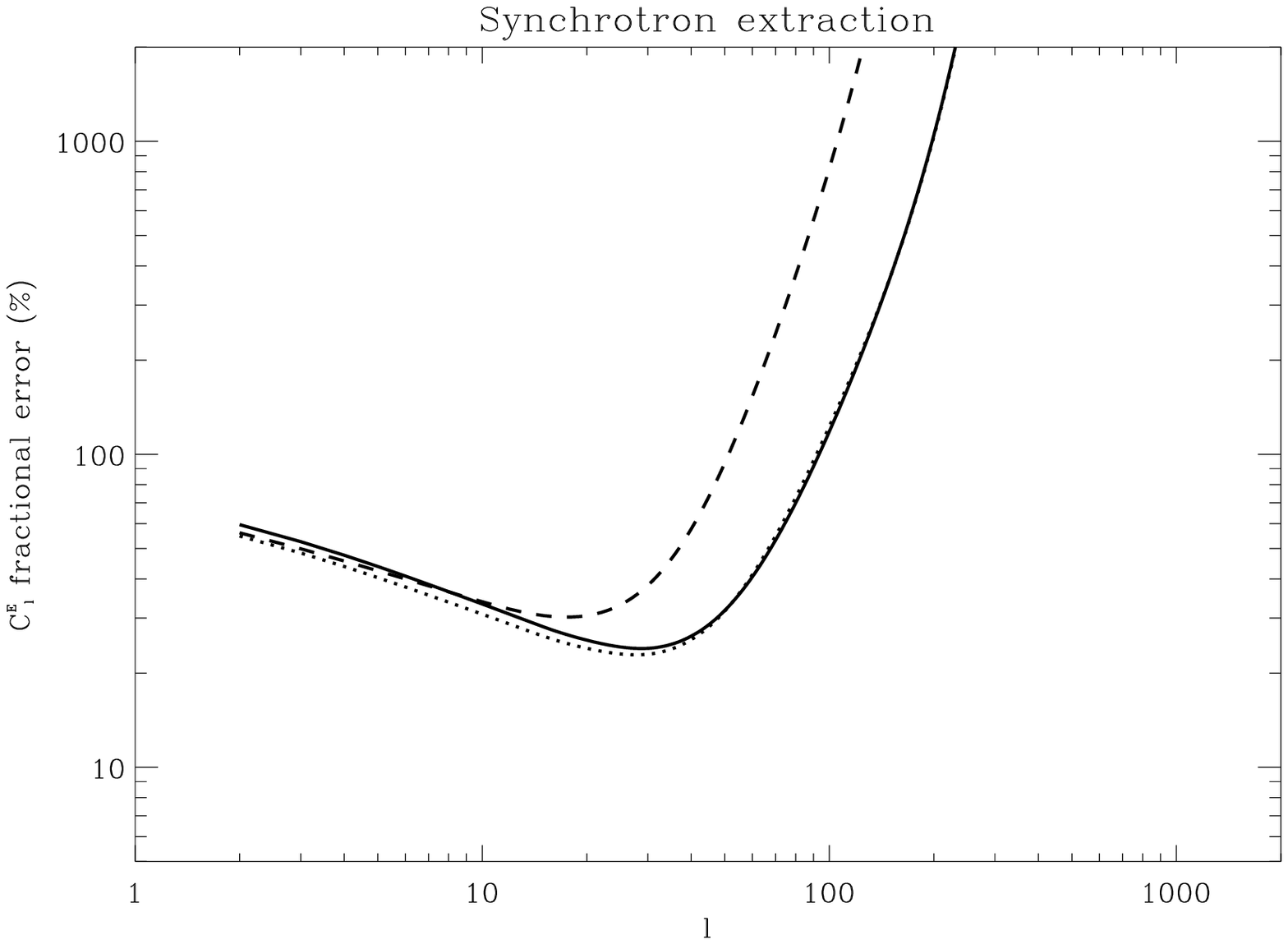,width=6cm,height=5cm}
\centering\epsfig{figure=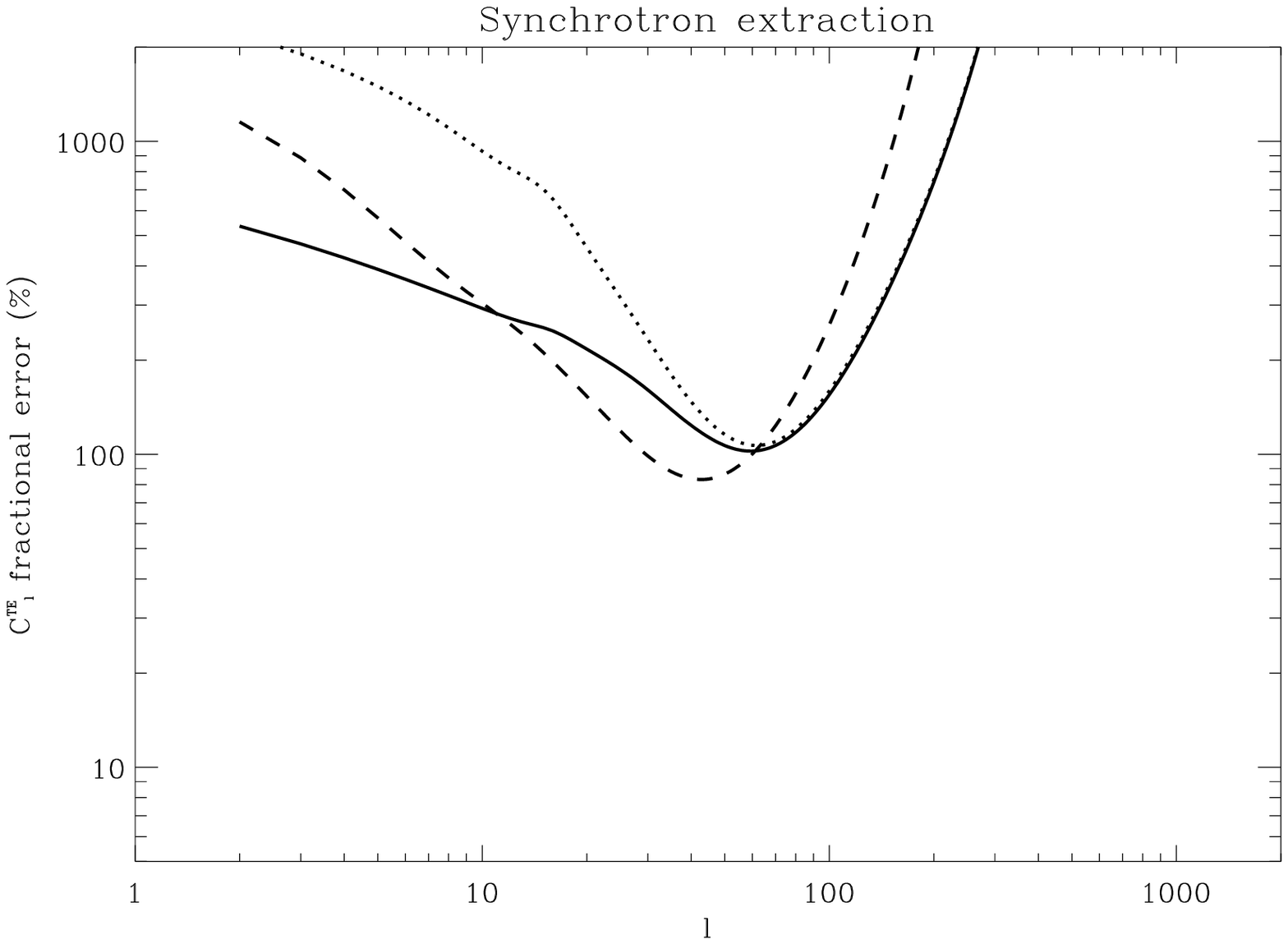,width=6cm,height=5cm}
\caption{Extraction of synchrotron $E$ (Left Panel)
  and $ET$ (Right Panel) power spectra with LFI ({\em
dotted \/} line), full \plancks ({\em solid \/} line), and MAP ({\em dashed
\/} line).}
\label{cov_syn}
\end{figure}

As already pointed out in the discussion on quality factors, polarised
foregrounds from dust and synchrotron have a very good chance of being
extracted as well as the CMB signal from future experiments. The basic
requirement for extracting a process well is that it dominates the total
signal at  least one frequency of the experiment. Both the future
experiments MAP and \plancks will have frequency channels which fulfill this
condition---MAP's two lowest frequency channels, centered at $22$ and $30 \,
\rm GHz$, and LFI's $30 \, \rm GHz$ channel are likely to be dominated by
polarised synchrotron; the $545 \, \rm GHz$ channel of \plancks will act as a
template for polarised dust.  In Figs.~(\ref{cov_dust}) and (\ref{cov_syn}) we
plot the fractional error on the power spectra of various Galactic polarised
foregrounds. The polarised component of dust can be extracted even better than
the CMB using HFI because of its $545 \, \rm GHz$ channel. Similarly,  the
presence of the lowest frequency channel on MAP and LFI should enable one to
extract the synchrotron signal for $\ell \le 100$. It should be noticed that
for all the experiments the fractional errors on $ET$ cross-correlation are
larger than the $E$ power spectrum. It is because the $ET$ cross-correlation
of foregrounds is mixed with a relatively large  CMB $ET$ cross-correlation  
signal  and therefore does not dominate the total signal as the $E$-mode
 foreground polarisation.  Another noticeable feature of the 
figures is that MAP performs as well or even better than LFI and \plancks 
in extracting the polarised component of synchrotron. It is because  
MAP has the   lowest frequency channel at $22 \, \rm GHz$. Therefore,
despite lower sensitivity, MAP has better frequency coverage at
frequencies which are dominated by polarised synchrotron signal.

It should be pointed out that the precision of extraction of a given process
depends crucially on the other competing processes. For instance, the
unpolarised component of synchrotron cannot be extracted as well the
unpolarised component not only because the unpolarised component never
dominates the total signal at any frequency but also because it is very
difficult to extract it from a comparable level of the free-free
signal. Therefore, the quality of extraction of synchrotron polarisation shall
depend quite sensitively on the presence of other sources of polarised
foreground like polarised free-free emission or the extragalactic radio
sources which are seen to be polarised at $\la  20 \%$ level (Saikia \&
Salter 1988). The dust
polarised signal, however, is unlikely to be affected as there are no known
sources of polarised foregrounds in the high frequency range covered by HFI,
 except infra-red point sources which might be polarised but will affect the
signal only for $\ell \ge 1000$ (Toffolati {\em et al. \/} 1997).

\subsection{Detection of B-mode polarisation}

An unambiguous way to infer the presence of gravitational waves in the early
universe is through the detection of $B$-mode polarisation (Seljak 1997,
Kamionkowski \& Kosowsky 1997,  Seljak \&  Zaldarriaga 1997). (Gravitational
lensing can generate a $B$-mode signal from a purely $E$-mode primary input
. However, this signal is weak and peaks around $\ell \simeq 1000$
(Zaldarriaga \& Seljak 1998, Bernardeau 1998). We neglect it in this paper.)
This signal is much smaller than the $E$-mode polarisation and is negligible
for $\ell \ga 100$, but it is potentially detectable because of the high
sensitivity of \planck. We plot the fractional errors on this quantity in
Fig.~(\ref{cov_bmode}) for the specifications of \planck.  The base model is
tilted CDM with the scalar index $n_s = 0.9$, the tensor index $n_t = -0.1$,
and ratio of tensor to scalar quadrapole $T/S = 0.7$.  This signal is barely
above the noise level and is comparable to the foreground contamination
(Prunet {\em et al. \/} 1998).  However, Fig.~(\ref{cov_bmode}) suggests that
it might be possible to get a marginal detection with \plancks at low $\ell$.

\begin{figure}
\centering\epsfig{figure=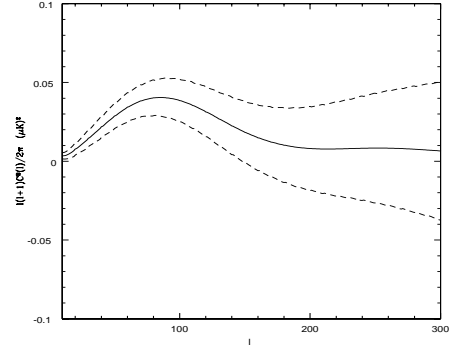,width=6cm,height=5cm}
\caption{The  CMB $B$-mode signal is plotted along with the expected
  $1\sigma$ measurements with band power taken with a logarithmic
  interval $\Delta \ell /\ell = 0.2$ for \plancks.}
\label{band_b}
\end{figure}

A noteworthy feature of the $B$-mode detection is that the Wiener
filtering result  is quite comparable
to the performance of the combined sensitivity of all the
channels. 
This is largely because the $B$-mode signal does not correlate with
any other signal and is therefore free of errors coming from 
cross-correlations with  foregrounds of other fields ($T$ and $E$). 
 Our results suggest that we expect the answer to lie between
 the case of best channel and the combined sensitivity of all the
 channels. With band-power estimates with a $20 \%$ logarithmic
 band, the errors
 drop  to $\le 25 \%$ for $ 20 \le \ell \simeq 100$. These
are shown in Fig.~(\ref{band_b}).

Also, as seen in Fig.~(\ref{cov_bmode}), the $B$-mode dust polarisation can be
extracted with much better precision than the CMB signal. It is because though
the CMB $B$-mode signal is more than an order of magnitude
below the $E$-mode signal, we 
took them to  be comparable for foregrounds.
 This would mean that the $B$-mode foregrounds  will 
dominate the signal at most of the frequencies of the future experiments, and 
therefore can be extracted better than the CMB signal.  The synchrotron
$B$-mode signal, however, is seen to be much harder to extract.  

\section{Conclusion and discussion}

We devised a multi-frequency Wiener filtering method to consider the effect of
Galactic polarised foregrounds on the detection of CMB $E$-mode polarisation
and $ET$ cross-correlation using future CMB missions. Our results can be
summarized as:

\begin{enumerate}
  
\item The foregrounds can be subtracted well enough for the LFI and HFI
    aboard \plancks to detect the $E$ and $ET$ signal with signal-to-noise
    $\simeq 2 \hbox{--} 10$ for most of the $\ell$-range in $50 \le
    \ell \la 1000$.
  
\item The foregrounds are likely to greatly undermine the performance
    of MAP. It seems unlikely that MAP could detect either $E$ or $ET$ signal
    for individual modes. However, by taking band-power estimates with a
    $20\%$ logarithmic interval, noise levels reduce sufficiently to allow a
    marginal detection of the $ET$ signal 
    near $\ell \simeq 300$.
    
\item The power spectra of $E$-mode  polarised dust can be extracted for  $\ell
  \la 1000$ range using \plancks HFI with signal-to-noise
  between 1 and 10. The $ET$ cross-correlation of this contaminant can
  be detected with signal-to-noise $\le 2$ for $ 100 \le \ell \le 1000$. 
   The $E$-mode  power spectrum of
  polarised synchrotron, on the other hand, can only
   be determined by either  MAP and \plancks LFI
    for $\ell \le 100$.  This  suggests that {\em both MAP and 
      \plancks } {\em  have  a fairly good chance of determining}
    {\em the polarised foregrounds which are}
    {\em expected to hamper their  performances in the detection of the CMB}
    {\em polarisation, at least for a small range of $\ell$.}  

\item The $B$-mode polarisation, which
    unambiguously establishes the presence of stochastic gravitational waves
    from the inflationary era, can be detected with signal-to-noise $\simeq 1$
    by \plancks for $\ell \la 100$. However, band-power estimates with a $20\%$
    logarithmic band will enable its detection with signal-to-noise $\simeq 2
    \hbox{--}4$ for $20 \le \ell \le 100$.

\end{enumerate}
    
\begin{figure}
\centering\epsfig{figure=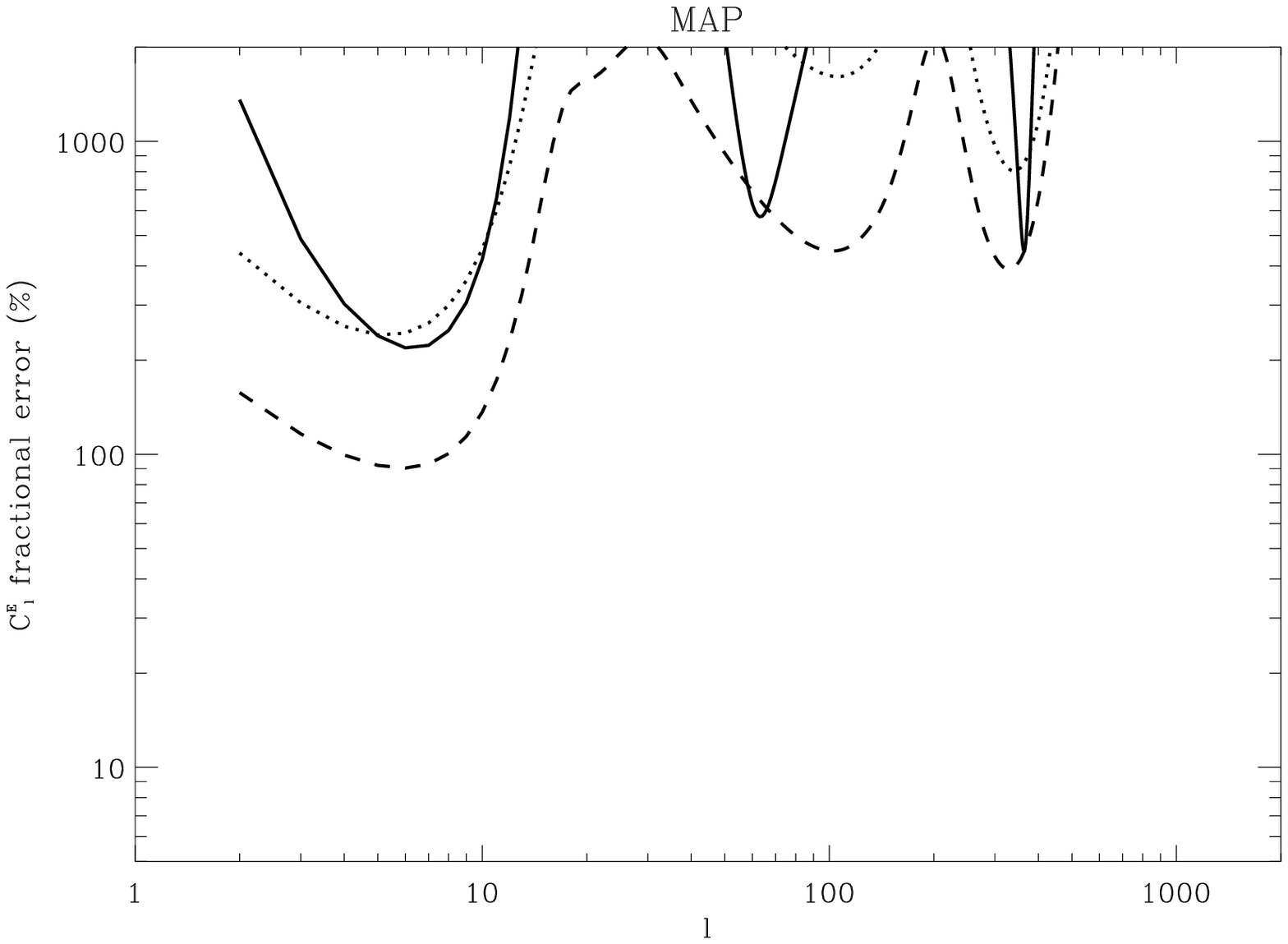,width=6cm,height=5cm}
\centering\epsfig{figure=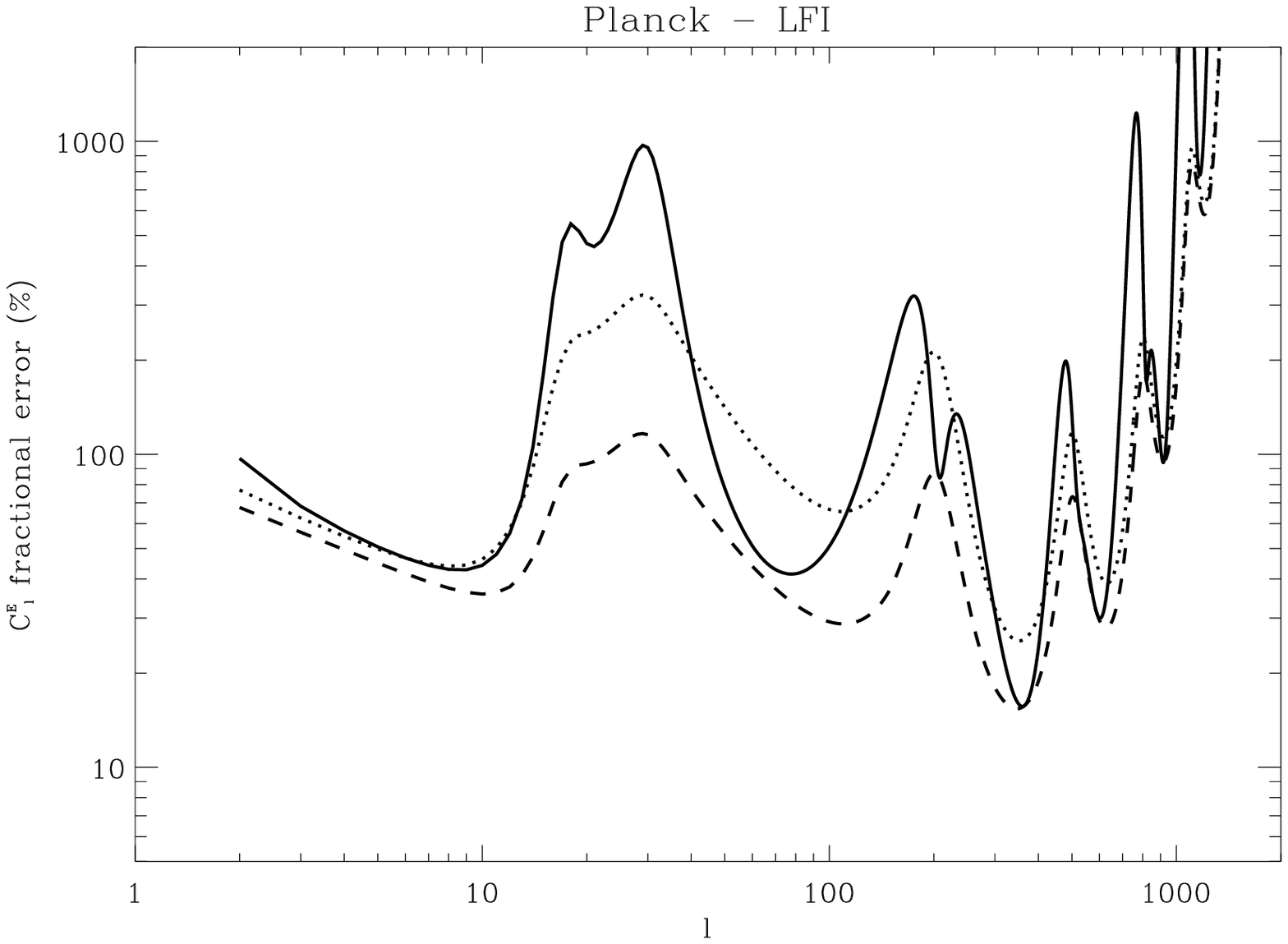,width=6cm,height=5cm}
\centering\epsfig{figure=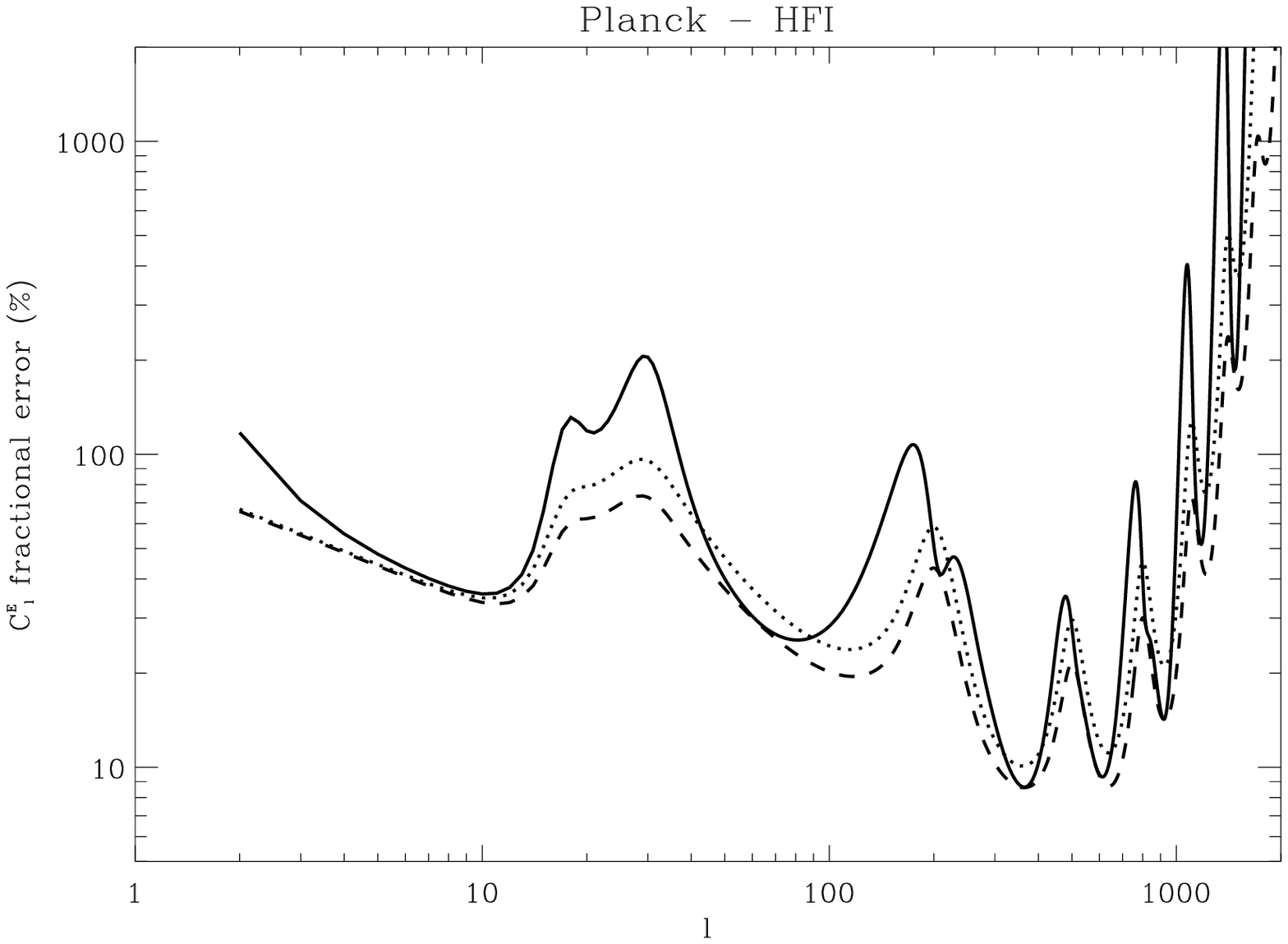,width=6cm,height=5cm}
\centering\epsfig{figure=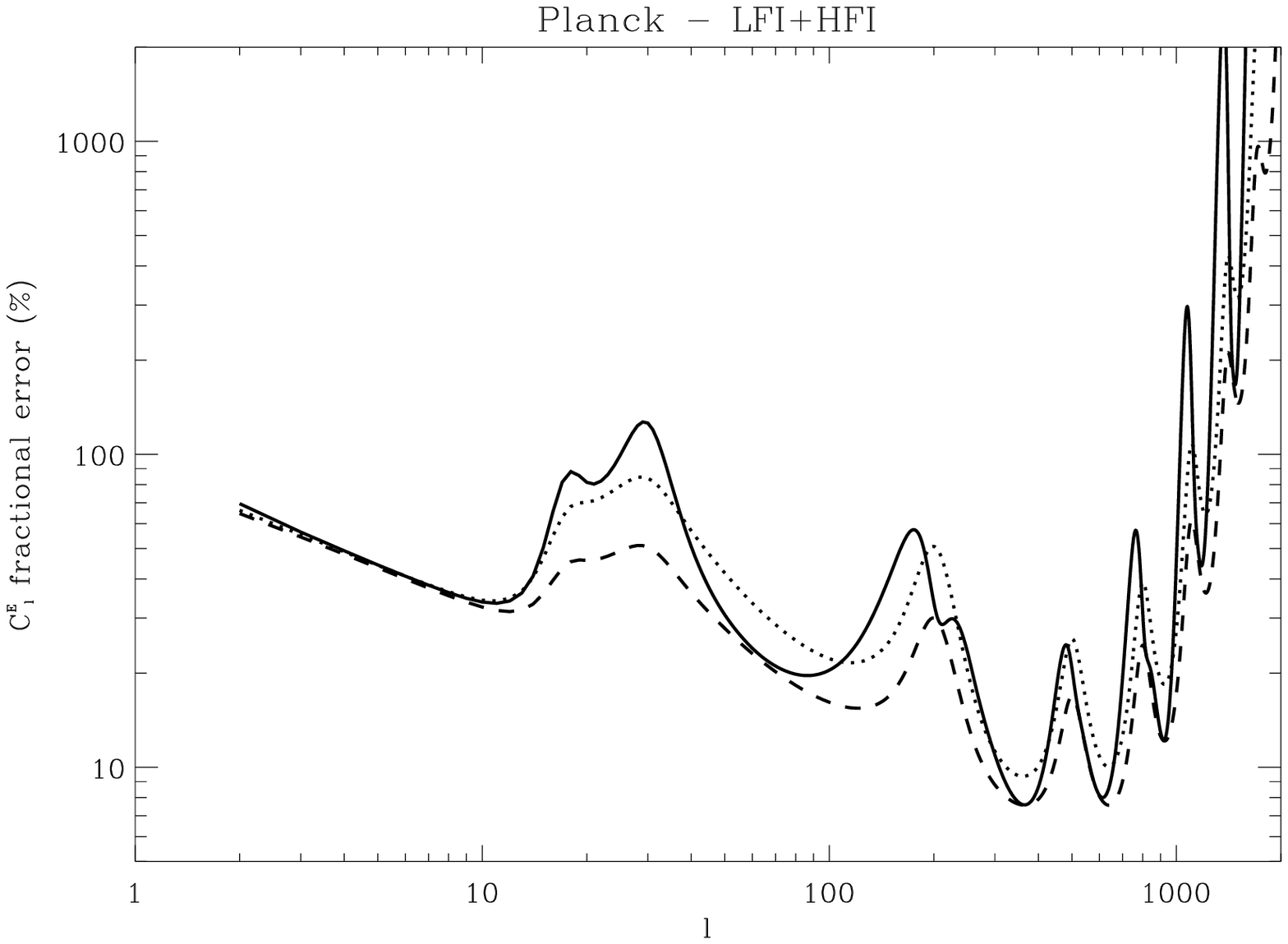,width=6cm,height=5cm}
\caption{Fractional errors in extracting the $E$ mode polarisation. The {\em
solid}, {\em dotted}, and {\em dashed \/} lines correspond to the expected
performances of Wiener filtering, best channel, and combined sensitivity of
all the channels of a given experiment, respectively.}
\label{cov_pol}
\end{figure}

\begin{figure}
\centering\epsfig{figure=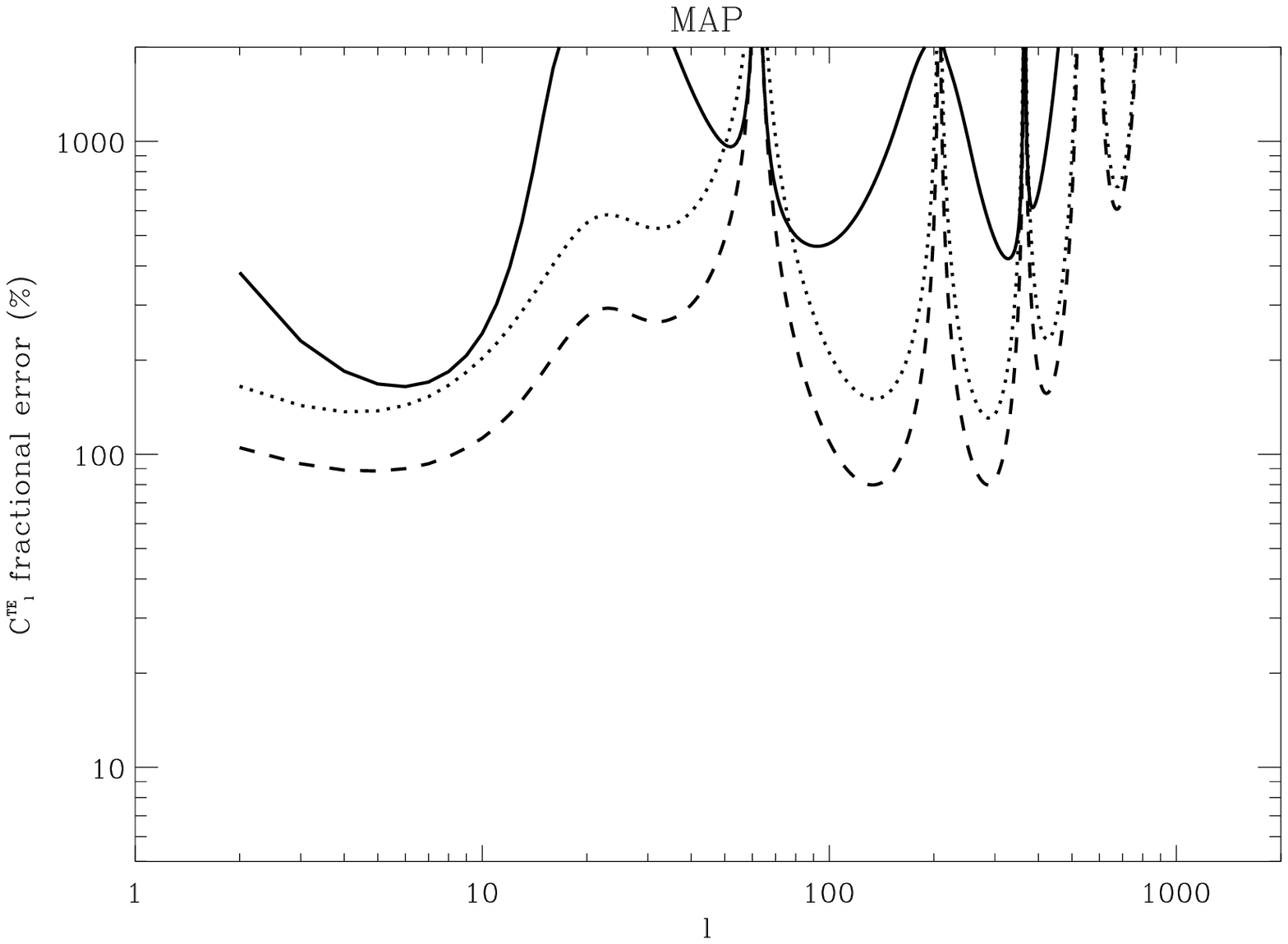,width=6cm,height=5cm}
\centering\epsfig{figure=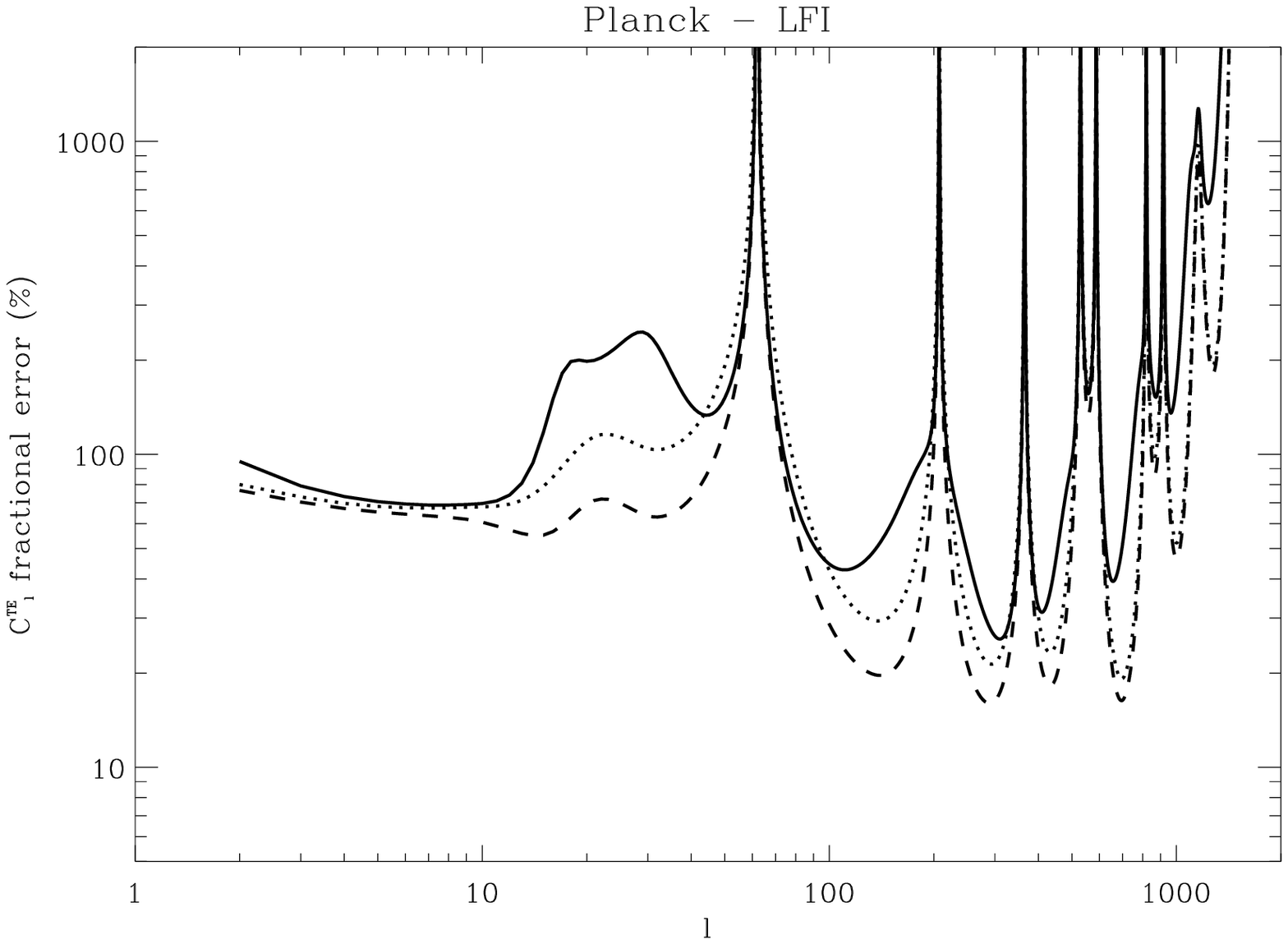,width=6cm,height=5cm}
\centering\epsfig{figure=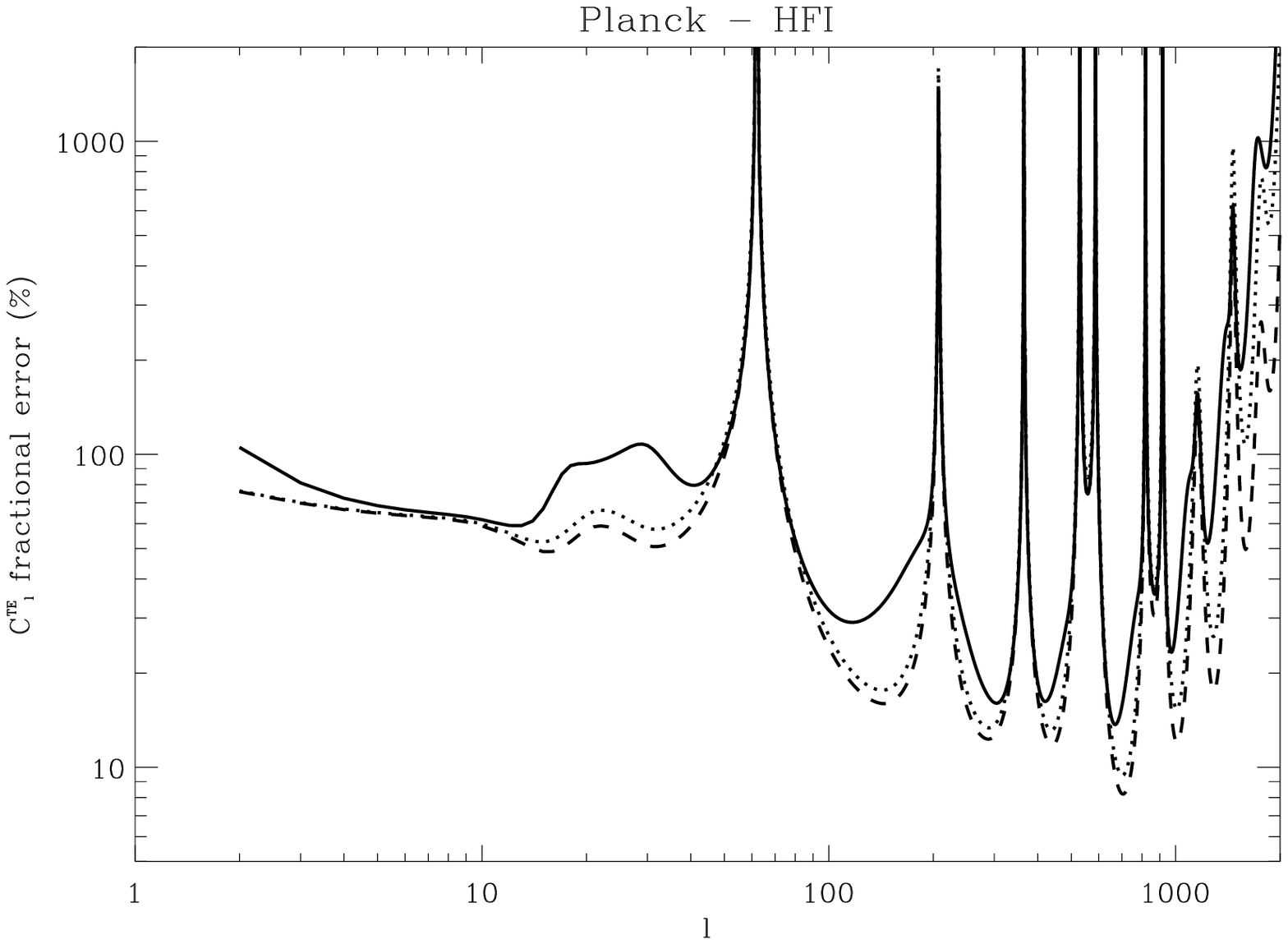,width=6cm,height=5cm}
\centering\epsfig{figure=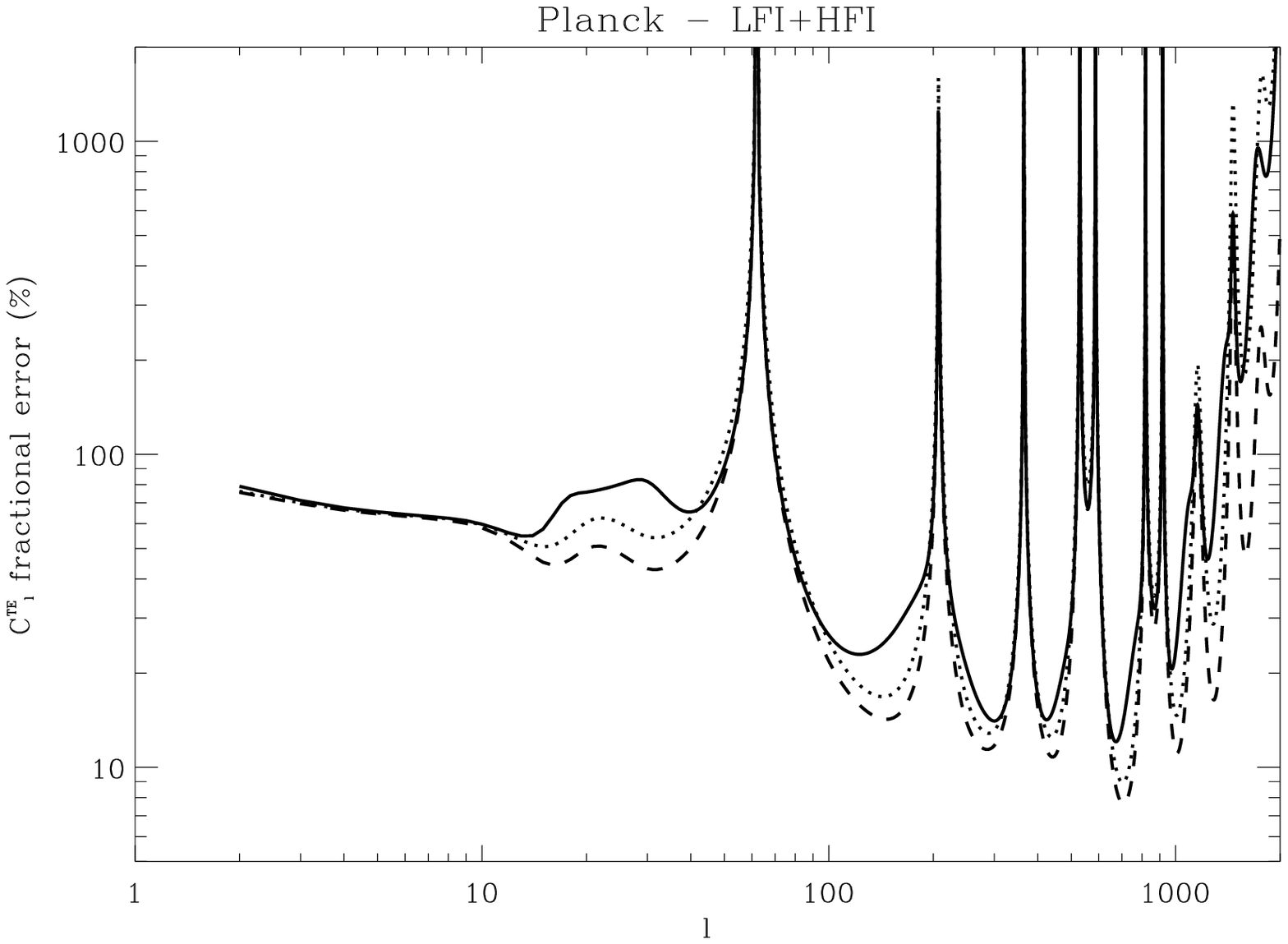,width=6cm,height=5cm}
\caption{Same as Fig.~\ref{cov_pol} for the $ET$
  cross-correlation. The sharp spikes in the figure merely indicate
  the values of $\ell$ at which the signal vanishes.}
\label{cov_cross}
\end{figure}

\begin{figure}
\centering\epsfig{figure=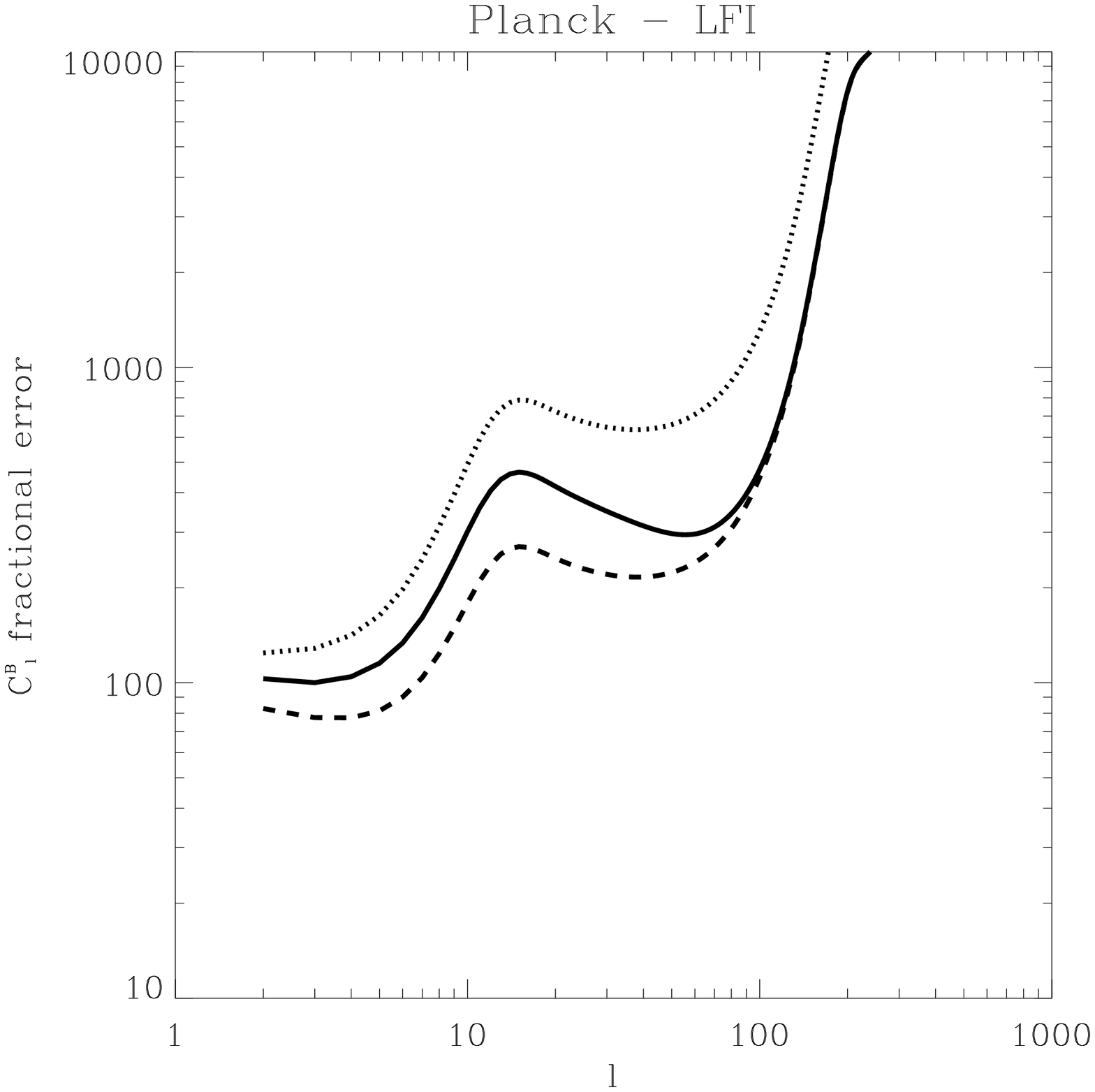,width=6cm,height=5cm}
\centering\epsfig{figure=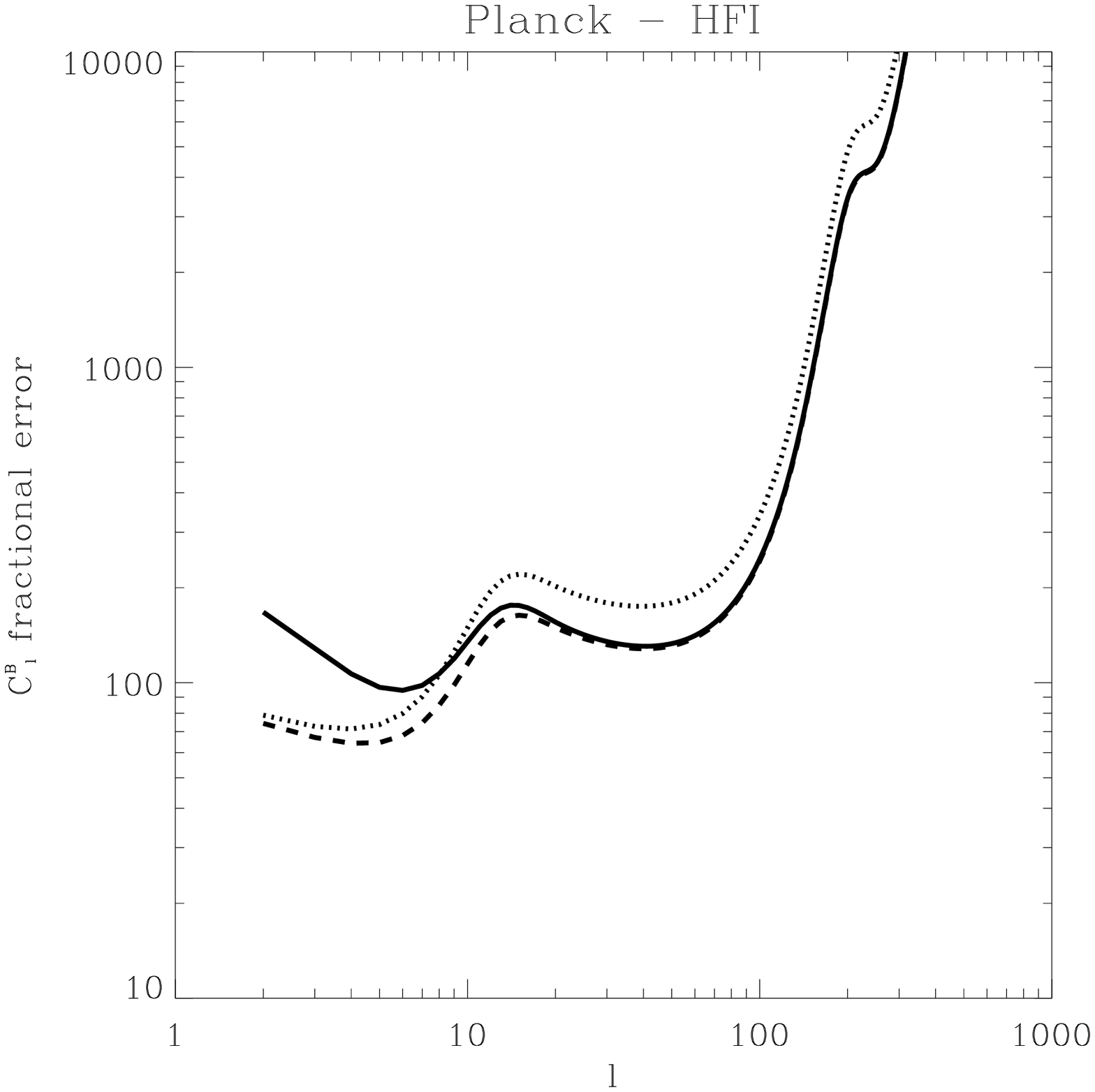,width=6cm,height=5cm}
\centering\epsfig{figure=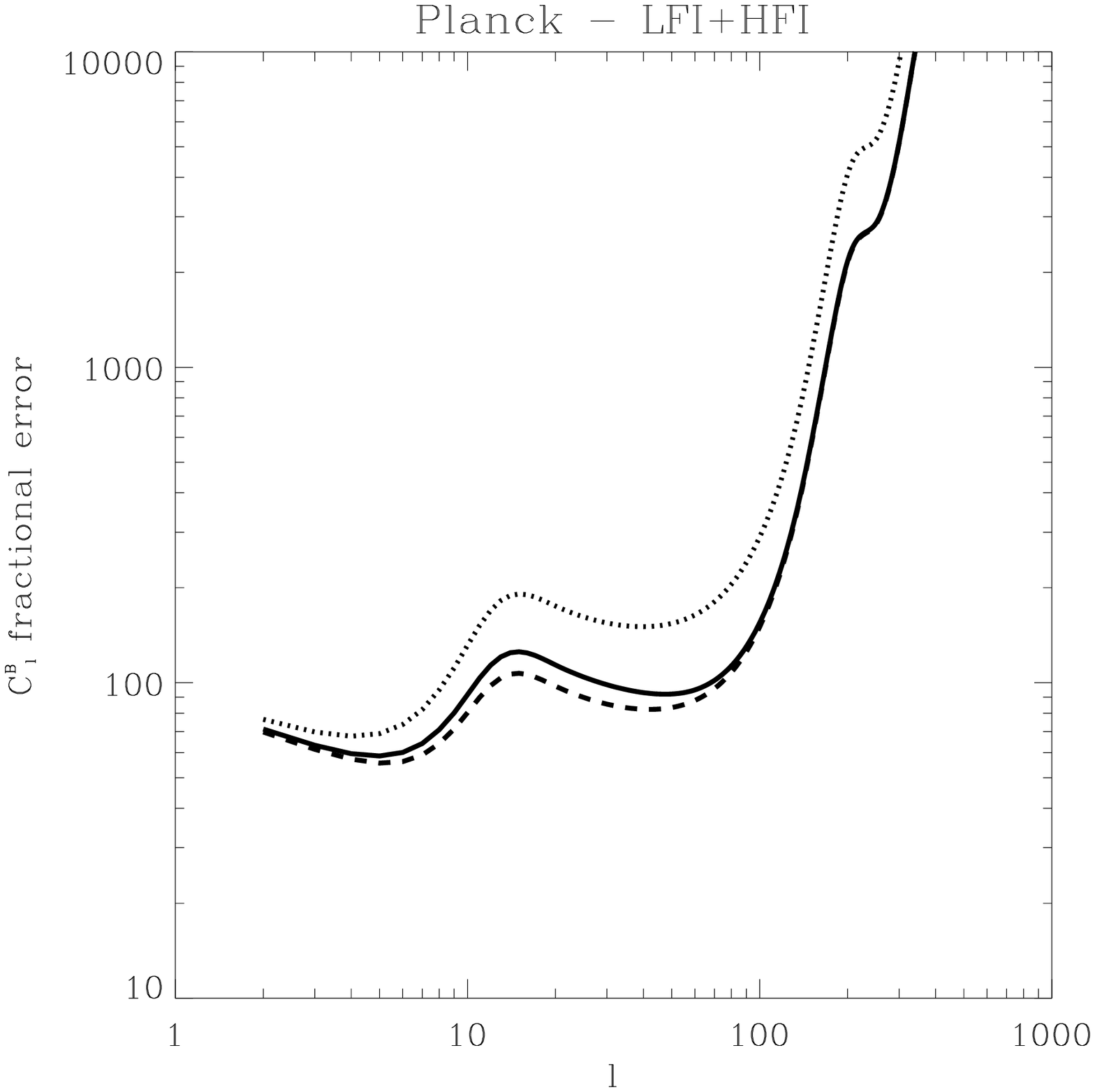,width=6cm,height=5cm}
\centering\epsfig{figure=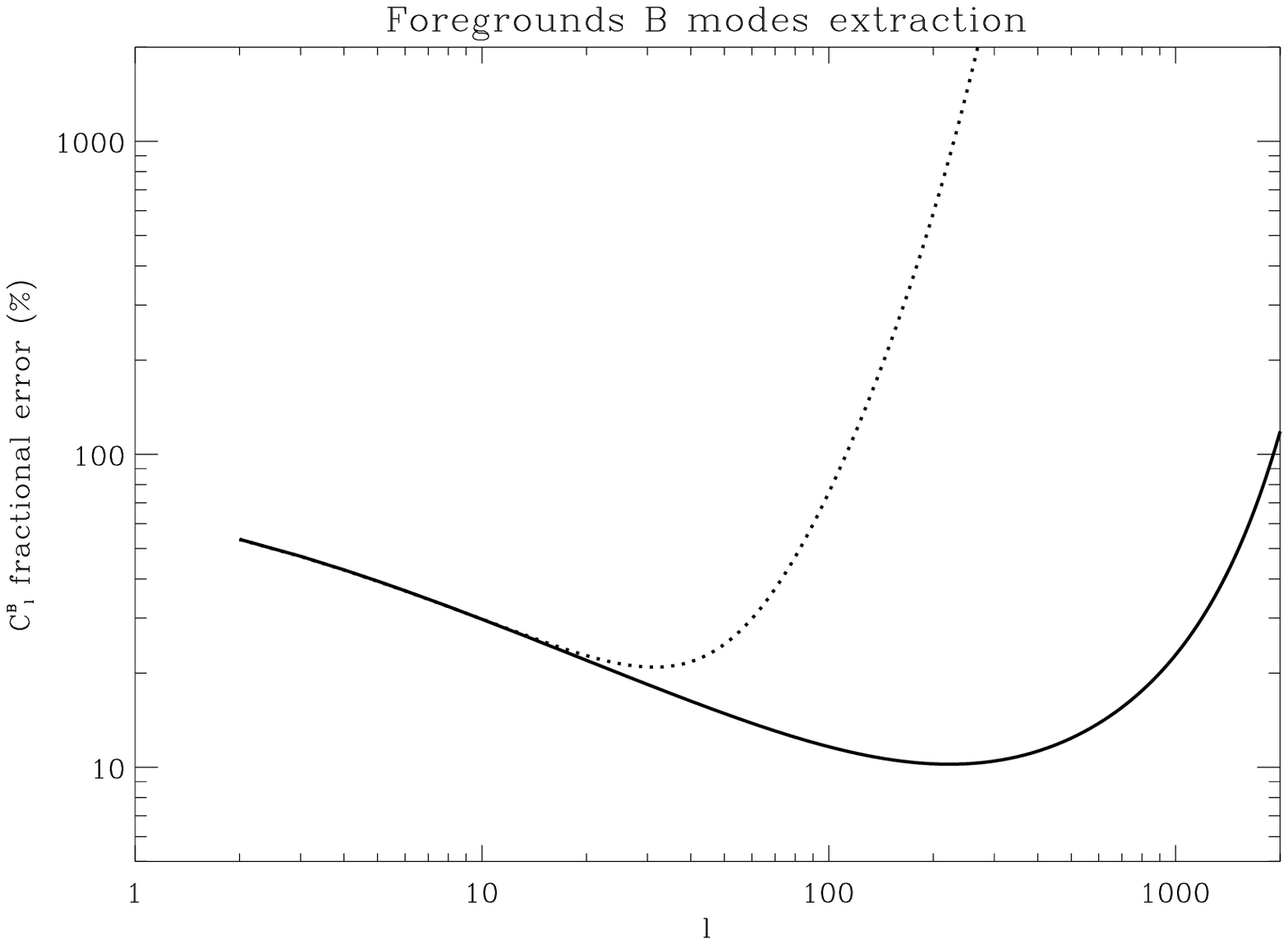,width=6cm,height=5cm}
\caption{Fractional errors for $B$-mode polarisation are shown in the first
  three panels. The {\em solid \/}, {\em dotted \/}, and {\em dashed \/}
 curves correspond to Wiener, best channel, and combined sensitivity,
 respectively. The forth panel shows the corresponding errors
 in the extraction of the $B$-mode component of the polarised dust
 ({\em solid \/} line)  and synchrotron ({\em dotted \/} line). The
underlying model was taken to be a CDM model with $n_s = 0.9$,
$T/S = 0.7$,  and $n_T = 0.1$} 
\label{cov_bmode}
\end{figure}

Another possible way to detect very
small signals ($E$ and $ET$ signal with MAP and
$B$ signal with \planck) is to image a small fraction of the sky at high
Galactic latitudes for longer periods.  It is expected that \plancks will
image around $1\%$ of the sky at high Galactic latitudes for periods
$5\hbox{--}6$ times the all sky average. One advantage of this approach is
that foreground level at high Galactic latitudes is much smaller than the all
sky average (e.g. dust emission is smaller by nearly a factor of 10 at
Galactic latitudes north of $70^\circ$ as compared to the all sky average we
use in this paper). However, this will increase the covariance of power
spectra by a factor of $\sqrt{f_{\rm sky}}$. In light of our results one could
ask whether it is preferable to deep-image a part of the sky with small
foregrounds or one should decrease the covariance by covering a larger
fraction of the sky, albeit with higher foreground levels. It is seen from
Figs~\ref{cov_pol} and \ref{cov_cross} that the expected errors on the power
spectra of $ET$ and $E$ are, for $\ell \simeq 100$, a factor of 6 more than
the best performance of MAP without foregrounds. It might be possible in this
case to gain at least a factor of 2 (in addition to sensitivity gained by
decrease in pixel noise from longer integration time) by imaging $10\%$ of the
sky at higher Galactic latitudes where the foreground level is expected to be
much smaller. However, this is not the case for the detection of $B$-mode
polarisation by \planck. As seen in Fig~\ref{cov_bmode}, Wiener filtering
extracts the signal almost as well as one could get using the combined
sensitivity of all the channels. This means that if $1\%$ of the sky is
integrated for a period $\simeq 5$ times more than the all sky average, the
fractional error will only {\em increase} by a factor of $\simeq 4$ because a
decrease in foreground level will not affect the fractional errors.

In going from Eqs.~(\ref{unbiased}) to Eqs.~(\ref{covar1}), (\ref{covar2}), and
(\ref{covar3}), we assumed the foregrounds to be Gaussian, i.e., the 4-point
correlation function was assumed to be expressible as a combination of 2-point functions
only. This assumption is  erroneous for the foregrounds. The
irreducible 4-point function from foregrounds can add to the covariances
thereby enhancing the errors.  We cannot quantify this increase within the
framework of this work. However, as pointed out above some of the frequency 
channels in the future experiments will be dominated by foregrounds. A direct 
analysis of these maps is likely to reveal the non-Gaussian nature of 
foregrounds, which could then be used to quantify errors on CMB extraction.

{\em Dependence on input model \/}: The results presented in the paper 
obviously depend on the choice of  model chosen for generating
the CMB fluctuations. We used a variant of CDM
model with reionization for the $E$ and $ET$ extractions. Reionization 
suppresses anisotropies at small angular scales by $\exp(-2\tau)$, which is
not a major effect for $\tau = 0.1$ that we choose. The most important effect
of reionization is to generate new polarisation anisotropies at
 $\ell \simeq  2\hbox{--}30$ depending on the optical depth. In the absence of 
 reionization, the signal will be smaller by more than an order of
 magnitude
 for $\ell \la
15$.  Therefore, the validity of  our results   will be very
sensitive to the underlying model for $\ell \le 15$.

The detection of polarisation
anisotropies at smaller scales should not depend so much on 
the choice of model, because  most variants of sCDM  models
give comparable level of these 
anisotropies for  $\ell \ge 20$. 
It should be true unless the optical depth to the large scattering surface 
is  large ($\tau \simeq 1$) or that the large scale anisotropies 
are dominated by tensor anisotropies. However, the current data 
for temperature anisotropies already  suggest that the first 
Doppler peak is even  higher than predicted by sCDM model
 (Netterfield {\em et al. \/} 1997), which rules out the
possibility of strong reionization and makes it  difficult for the 
scalar anisotropies to be sub-dominant to contribution from gravitational 
waves.  Therefore, it is safe to conclude that our predictions for
$\ell \ge 20$ will not be seriously affected by a change in the underlying
model.  

For $B$-mode polarisation we considered a model with tensor to scalar quadruple
ratio $T/S = 0.7$. As mentioned above, this will lower the contribution from
scalar modes and therefore will reduce the 
signal-to-noise for  the  detection of $E$-mode 
polarisation. However, the 
$B$-mode signal is roughly proportional to the value of $T/S$ within the 
framework of inflationary models which require $T/S \simeq  -7n_T$.  As shown 
above, the $B$-mode foregrounds can be subtracted quite efficiently and for
this model the 
signal-to-noise for CMB $B$-mode signal   is $\simeq 2 \hbox{--} 4$. 
Therefore, it might be possible to detect a signal with a value as
small as  $T/S \simeq 0.2 \hbox{--} 0.3$. 

The Wiener filtering method assumes
{\em a priori \/ } knowledge of the power spectra of CMB and foregrounds as
well as the frequency dependence of foregrounds. Therefore, the error of 
extraction using this method does not include the error in
evaluating these input quantities. Future  experiments will make 
multi-frequency maps
with millions of pixels. This will make it  difficult to apply the usual
maximum likelihood technique to extract various power spectra. Fast
methods for tackling this problem are currently being developed (Oh
{\em et al. \/} 1998). In future, it should become feasible to
quantify  errors on the extraction of the power spectra from high
resolution multiple frequency maps. It should then be  possible to
revise our estimates of the errors.


\noindent

\noindent

\noindent

\noindent

\noindent

\noindent

\noindent

\noindent

\noindent

\noindent

\noindent

\noindent

\end{document}